\documentclass[twocolumn,tighten]{aastex62}
\pdfoutput=1 %for arXiv submission
\usepackage{amsmath,amstext}
\usepackage[T1]{fontenc}
\usepackage{ae,aecompl}
\usepackage[utf8x]{inputenc}
\usepackage{atbegshi}
\usepackage{apjfonts} 
\usepackage{graphics}
\usepackage[figure,figure*]{hypcap}
\usepackage{enumitem}
\usepackage{url}
\usepackage{bm}
\usepackage{lineno}
\usepackage[normalem]{ulem}
\input{hyperlink-year-only-natbib-patch}

\usepackage{lipsum}
\graphicspath{{./}{figures/}}

\usepackage[T1]{fontenc} % if needed

\reportnum{DES-2020-0559}
\reportnum{FERMILAB-PUB-20-334-AE}

\begin{document}

%\mar{DES-2020-0559}

\title[splashback]{Probing galaxy evolution in massive clusters using ACT and DES: splashback as a cosmic clock}

%\correspondingauthor{Susmita Adhikari}
%\email{susmita@stanford.edu}
%\correspondingauthor{Tae Hyeon-Shin}
%\email{taeshin@upenn.edu}

\author[0000-0002-0298-4432]{Susmita Adhikari}
\email{susmita@stanford.edu}
\affiliation{Kavli Institute for Particle Astrophysics and Cosmology and Department of Physics, Stanford University, Stanford, CA 94305, USA}
\affiliation{SLAC National Accelerator Laboratory, Menlo Park, CA 94025, USA}
\author[0000-0002-6389-5409]{Tae-hyeon Shin}
%\correspondingauthor{Tae Hyeon-Shin}
\affiliation{Department of Physics and Astronomy, University of Pennsylvania, 209 S. 33rd St. Philadelphia, PA 19104, USA}
\email{taeshin@upenn.edu}
\author[0000-0002-8220-3973]{Bhuvnesh Jain}
\affiliation{Department of Physics and Astronomy, University of Pennsylvania, 209 S. 33rd St. Philadelphia, PA 19104, USA}
\author[0000-0002-6792-2208]{Matt Hilton}
\affiliation{Astrophysics Research Centre, University of KwaZulu-Natal, Westville Campus, Durban 4041, South Africa}
\affiliation{School of Mathematics, Statistics \& Computer Science, University of KwaZulu-Natal, Westville Campus, Durban 4041, South Africa}
\author[0000-0002-6836-3196]{Eric Baxter}
\affiliation{Institute for Astronomy, University of Hawaii, Manoa, Hawaii, USA}
\author[0000-0002-7887-0896]{Chihway Chang}
\affiliation{Kavli Institute for Cosmological Physics, The University of Chicago,
	Chicago, IL 60637, USA}
\affiliation{Department of Physics, The University of Chicago, Chicago, IL 60637, USA}
\author[0000-0003-2229-011X]{Risa H. Wechsler}
\affiliation{Kavli Institute for Particle Astrophysics and Cosmology and Department of Physics, Stanford University, Stanford, CA 94305, USA}
\affiliation{SLAC National Accelerator Laboratory, Menlo Park, CA 94025, USA}
\author[[0000-0001-5846-0411]{Nick Battaglia}
\affiliation{Department of Astronomy, Cornell University, Ithaca, NY 14853, USA}
\author{J. Richard Bond}
\affiliation{Canadian Institute for Theoretical Astrophysics, 60 St. George Street,
University of Toronto, Toronto, ON, M5S 3H8, Canada}
\author{Sebastian Bocquet}
\affiliation{Faculty of Physics, Ludwig-Maximilians-Universit\"at, Scheinerstr. 1, 81679 Munich, Germany}
% ...
\collaboration{(DES Collaboration)}
\collaboration{(ACT Collaboration)}

\begin{abstract}

We measure the projected number density profiles of galaxies and the splashback feature in clusters selected by the Sunyaev--Zeldovich (SZ) effect from the Advanced Atacama Cosmology Telescope (AdvACT) survey using galaxies observed by the Dark Energy Survey (DES). The splashback radius for the complete galaxy sample is consistent with theoretical measurements from CDM-only simulations, and is located at $2.4^{+0.3}_{-0.4}$ Mpc $h^{-1}$. We split the sample based on galaxy color and find significant differences in the profile shapes. Red galaxies and those in the green valley show a splashback-like minimum in their slope profile consistent with theoretical predictions, while the bluest galaxies show a weak feature that appears at a smaller radius. We develop a mapping of galaxies to subhalos in $N$-body simulations by splitting subhalos based on infall time onto the cluster halos. We find that the location of the steepest slope and differences in the shapes of the profiles can be mapped to differences in the average time of infall of galaxies of different colors. The minima of the slope in the galaxy profiles trace a discontinuity in the phase space of dark matter halos. By relating spatial profiles to infall time for galaxies of different colours, we can use splashback as a clock to understand galaxy quenching. We find that red galaxies have on average been in their clusters for over $3.2 ~\rm Gyrs$, green galaxies about $2.2 ~\rm Gyrs$, while blue galaxies have been accreted most recently and have not reached apocenter. Using the information from the complete radial profiles, we fit a simple quenching model and find that the onset of galaxy quenching in clusters occurs after a delay of about a gigayear, and that galaxies quench rapidly thereafter with an exponential timescale of $0.6$ Gyr. 
\vspace{1cm}
\end{abstract}

\section{Introduction}
\label{intro} 
%\linenumbers

Galaxy clusters, the largest bound objects in the universe, are a unique laboratory for studying the non-linear evolution of the universe. These objects can be observed across the electromagnetic spectrum in $X$-ray, submillimeter, and optical, with each band illuminating different components of the cluster. Large-scale surveys of the sky have provided us with statistical samples of clusters that give us the opportunity to study the co-evolution of galaxies and matter in these extreme environments in great detail. Understanding the evolution of the galaxy populations that make up a cluster and the underlying dark matter potential can provide fundamental insights into the connection between dark matter and galaxies.

The properties of galaxies in a cluster are significantly different from those in the field. One striking feature of clusters is that they appear to be dominated by red galaxies, with little or no star formation \citep{Oemler:1974yw,Dressler:1980wq, Dressler:1984kh, Balogh:1997bw,   Poggianti:1999xh}.  Intra-cluster processes like ram-pressure stripping \citep{GunnGott72, Abadi:1999qy}, strangulation \citep{Larson:1980mv}, harassment, or tidal disruption are known to quench star formation in galaxies as they orbit within the cluster. Understanding the response of baryonic matter to the dark matter halo environment and the evolution of star formation has been a long standing question in galaxy evolution and cosmology. The assembly and evolution of galaxies is expected to be closely related to the evolution of its parent dark matter halo \citep{Bullock:2001pf, Wechsler2002, Moustakas:2001ew, Conroy:2005aq,Cooray:2005mm,  Conroy:2008dx,Kravtsov:2012zs, Somerville:2014ika,  Wechsler:2018pic}. While the total spatial distribution of galaxies in a cluster halo is determined by the dark matter potential, studying spatial correlations between various galaxy properties is a powerful way to infer the co-evolution of dark matter and galaxies.

In recent years, detailed studies of the galaxy distribution within clusters have emerged as a robust way to gain insight into dynamic properties of the halo. In particular, examining the logarithmic slope of the cluster halo profile highlights a novel feature in the density profile termed the \textit{splashback} radius, where the slope reaches a minimum in a narrow localized region. The splashback radius relates to the splashback surface, which can be thought of as the physical boundary of a halo \citep{Diemer:2014xya, Adhikari:2014lna, More:2015ufa, Shi:2016lwp, Mansfield:2017}.
The splashback surface traces the location of the apocenters in the orbits of the most recently accreted matter onto a dark matter halo \citep{Adhikari:2014lna, Diemer:2017ecy, Diemer:2017uwt}. This boundary therefore separates the multi-streaming or ``virialized'' region of a halo, where particles or satellite galaxies are in orbits, from the region where there are only infalling galaxies or particles (see also \citealt{Aung:2020czp}). The splashback surface denotes a sharp boundary in phase space, and it has been found that the slope of the spherically averaged, logarithmic matter density profile reaches a minimum at this boundary. This can be explained by the fact that the density in the virialized region of the halo follows a Navarro Frenk and White (NFW)-like profile that approaches a slope of $~-3$ in the outer regions \citep{Navarro:1995iw}, while the density of particles in the infall region follows a power law with index $\sim -1.5$ \citep{Baxter:2017csy}; a sharp transition between the virialized and infalling regimes at splashback therefore causes the slope to reach a minimum in a narrow, localized region before it asymptotes back to the background value.  

The splashback radius, as a feature in the spherically averaged density profile of both matter and galaxies around dark matter halos, can be accessed observationally. In several recent studies \citep{More:2016vgs, Baxter:2017csy, Chang:2017hjt, Shin19, Zuercher:2018prq, Murata:2020enz} the splashback radius was measured in the  projected number density of galaxies around massive clusters selected both optically and using the Sunyaev--Zeldovich \citep[SZ,][]{Sunyaev:1972eq} effect. The splashback radius has also been measured in the dark matter distribution itself, using weak lensing of galaxies around massive clusters; while \citet{Umetsu:2016cun} and \cite{Contigiani:2018qxn} have attempted to measure the feature using a small sample of massive $X$-ray selected clusters, \citet{Chang:2017hjt} have used a large, statistical sample of optically selected, RedMaPPer clusters \citep{Rykoff2014} in the Dark Energy Survey \citep[DES,][]{Abbott:2005bi} data. The latter finds evidence for a significant steepening of the density profile at the edge of galaxy clusters. \citet{Xhakaj:2019qln} find that future surveys like LSST \citep{Abell:2009aa} and Euclid \citep{Laureijs:2011gra} will measure the splashback radius through weak lensing with nearly $10\%$ precision.  In \cite{Tomooka:2020zjc}  a sharp boundary was also measured in the line-of-sight velocity dispersion of spectroscopic galaxies around clusters in SDSS \citep{sdssAbazajian:2008wr}. \citet{Okumura2017} and \citet{Okumura2018} have also explored theoretical predictions for the splashback radius in the observed velocity field.  All these studies provide evidence for the existence of a sharp transition around cluster boundaries, a feature that must exist in the presence of dark matter. Furthermore, studies have shown that the location of this feature can be sensitive to the nature of gravity and dark matter itself \citep{Adhikari:2018izo, Banerjee:2019bjp}. 

While the splashback radius is observed as a feature in the spatial distribution of matter and galaxies (and their velocities), it also has an inherent timescale associated with it: the time for a particle or galaxy to reach the apocenter of its first orbit.  In other words, dark matter particles or galaxies that form the splashback region have been inside the halo for approximately one orbital time, from accretion to first apocenter. Unlike dark matter particles, galaxy properties like star-formation rates (SFRs) and morphology can evolve significantly on similar timescales if the cluster environment plays a role in galaxy evolution.  Since the splashback feature  is closely related to the orbital time it is possible to use it as a clock to study different populations of galaxies and their time-evolving properties.

A galaxy within a cluster evolves from being blue and star-forming to red over the course of its multiple orbits through the halo.
Depending on the specific mechanism of quenching, the quenching timescale can vary. The longer a galaxy orbits within a halo the more likely it is to be affected by the intra-cluster medium and evolve in color, which in turn changes the radial profiles of galaxies binned in color. Thus by using the shapes of the profiles of galaxies of different colors within the splashback radius, we can learn about the quenching timescale. 
In the scenario that a population of galaxies has not reached apocenter (for example if we assume that all blue galaxies quench at or right after pericenter passage) we should not expect to see a true splashback feature in its density profile.  We show below that this applies to the bluest galaxies in our sample. 

In this paper we investigate the relation between galaxy evolution and density distributions within dark matter halos, focusing on information contained in the splashback feature. In particular, we study whether the location of the minimum in the profile's slope can provide additional insight into properties of galaxies with different SFRs.  We study how this information can help us relate the distribution of observed galaxies to the dark matter phase space. 

Our analysis relies on a catalog of galaxy clusters \citep{ACTcluster_inprep} that have been observed via their SZ signal in submillimeter wave maps from the Atacama Cosmology Telescope  \citep[ACT,][]{Aiola_2020, Naess:2020wgi}. SZ selection is essentially redshift independent, and is close to mass limited because of the small scatter in the relationship between cluster mass and SZ observable. Moreover, clusters selected with the SZ effect are known to suffer less from systematic effects such as line-of-sight projections and triaxiality than optically selected clusters \citep[e.g.][]{Shin19}. We study the distribution of galaxies around the SZ-selected clusters from ACT by cross-correlating the cluster positions with galaxies from DES data \citep{Abbott:2005bi}.

This paper is organized as follows. In section \ref{sec:splashback_sims} we study and review the relation between infall time, density profiles and splashback radius from $N$-body simulations. In section \ref{sec:obs}
we describe observations of splashback radius using galaxy clusters selected in SZ data, as a function of galaxy color; in section \ref{sec:quenching} we model the quenching of star-forming galaxies as a function of infall time to infer relevant timescales for cluster-mass halos, and summarize our conclusions in section \ref{sec:conclusion}.

\section{Phase space and infall time of subhalos and dark matter in simulations}
\label{sec:splashback_sims}

The collisionless dark matter particles that collapse gravitationally to form dark matter halos occupy a very specific region in the  phase space of halo-centric radius and radial velocity. Particles, as they fall into the halo, follow a  trajectory in this space as a function of time, forming what is known as, the ``multistreaming'' region of the halo. In other words, the multistreaming region is where particles are orbiting in the potential of the halo, and at any point in space there are multiple streams of particles, at different velocities, that have entered the halo at different times. The spherically averaged 3D density profile is the integral of the phase space distribution over the radial velocity. 
The dynamics of galaxies that fall into cluster halos largely follow the collisionless dark matter particles and can also be traced in this space. In this section we study the evolution of particles and substructure around cluster halos in phase space and infer how observed density profiles are composed of components accreted at different times.

\subsection{Simulations}
\label{subsection:sims}

We use two separate simulations to study halo evolution. To study dark matter particles we use a cosmological $N$-body simulation of Cold Dark Matter (CDM) in a $1 ~{\rm Gpc} ~h^{-1} $ box, with $1024^3$ particles, run using Gadget2 \citep{Gadget2}.  We study a sample of cluster halos selected based on their $M_{500c}$, the mass within the radius that encloses 500 times the critical density of the Universe. We apply a minimum mass threshold such that the mean mass of our sample  $\langle M_{\rm 500c}\rangle=3.1\times 10^{14} ~ M_{\odot} h^{-1}$. We extract all particles around the cluster halos at $z=0$ and find the orbits of sub-sampled sets of 1000 particles, randomly selected,  within $5 ~{\rm Mpc} ~h^{-1}$ of each object. To study the distribution of subhalos and halos around clusters we require simulations with higher resolution; for this we use the publicly available Multi-dark Planck \citep[MDPL2,][]{Multidark} simulation. This is a CDM-only simulation of a $1~{\rm Gpc^3}~h^{-3}$ volume with $3840^3$ particles. The simulation assumes the best-fitting flat $Lambda$CDM Planck cosmological model \citep{Ade:2013zuv}, with $\Omega_m=0.307$ and $h=0.677$. The properties of the halos and their subhalos were obtained using the {\sc Rockstar} halo finder \citep{Rockstar} in both the simulations. 

 Henceforth, we will refer to the host, cluster mass dark matter halos as ``clusters'' to avoid confusion while referring to the other halos around these cluster mass objects. This is primarily because we study the overall distribution of collapsed structures around clusters in the subsequent sections, that include both subhalos, within the virial radius, and other halos in the vicinity of the cluster extending beyond the virial radius. We use several timescales and lengthscales in this paper, a summary of variables is provided in Table \ref{tab:variables} for reference; we describe them in the text where they appear.

\begin{figure*}
	\centering
	\includegraphics[width=0.9\linewidth]{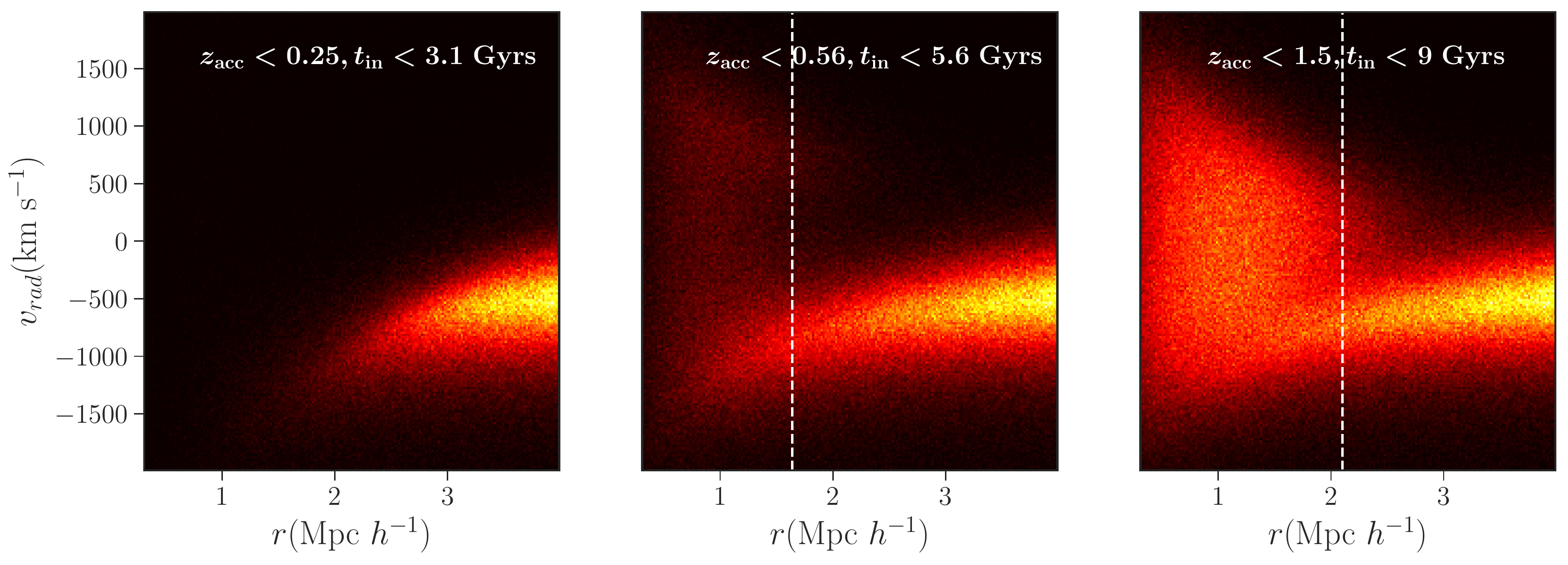}
	\includegraphics[width=0.7\linewidth]{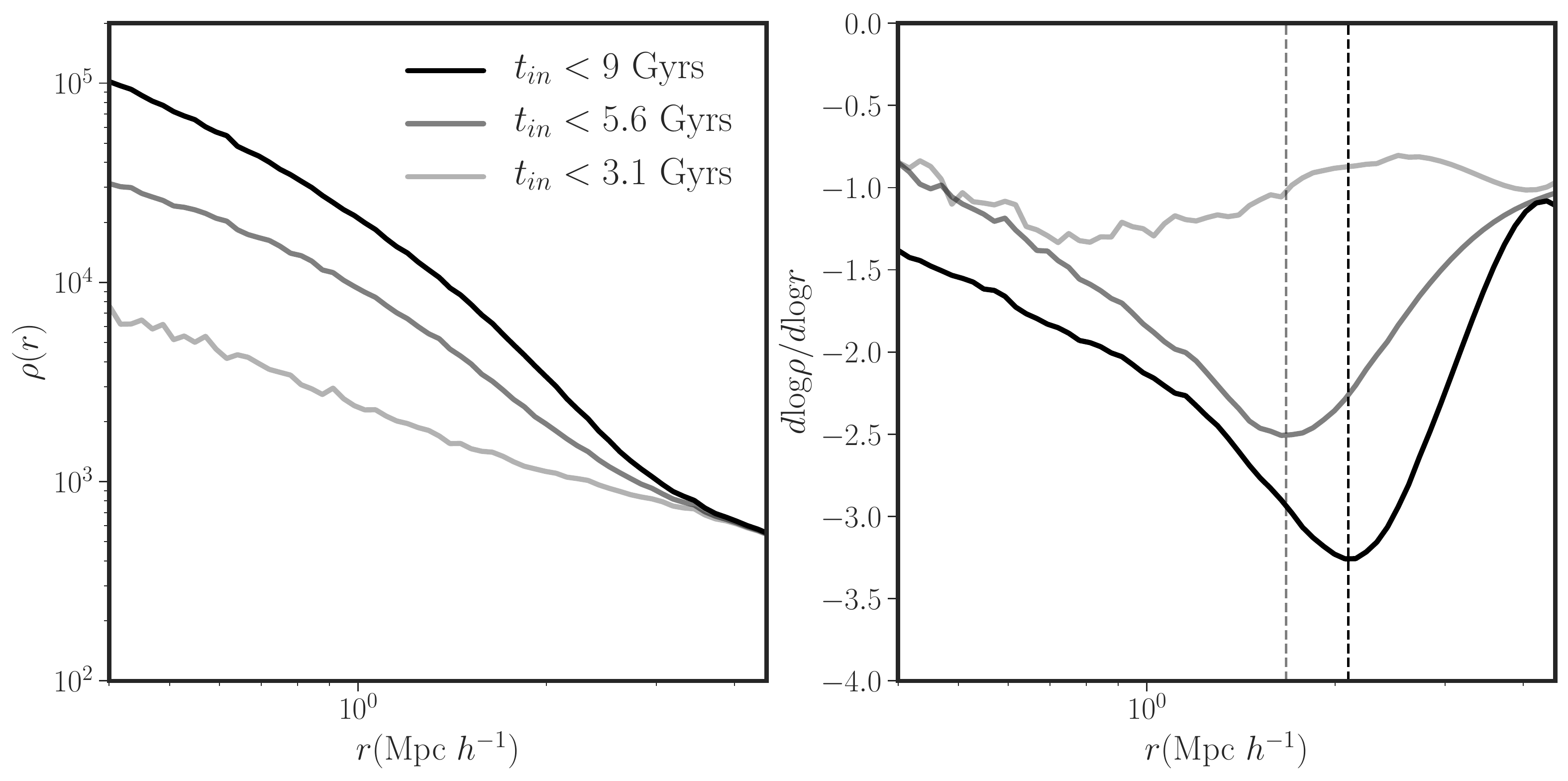}
	\caption{ (Top) Stacked phase space of particles accreted with different maximum time thresholds. Moving from large to small cluster-centric radius,  the left-most top panel corresponds to particles that have been within $4 ~{\rm Mpc}~ h^{-1}$ of the host for less than $3 ~\rm Gyrs$, the second and third panel correspond to particles that have been inside for less than $5.6$ and $9$ Gyrs respectively. The particles that have been inside for the least amount of time are present only in the infall stream (left panel). The middle panel shows a discontinuity (transition from multi-stream to infall stream), shown by the white vertical line, at a distance smaller than the traditional splashback radius. The actual splashback radius (right panel) is traced only by particles that have been inside the cluster long enough to reach apocenter. (Bottom) The left panel shows the density profiles for the three cases; the right panel shows the logarithmic slope. The location of the minimum in the slope profile, $r_{\rm smin}$, coincides with the location of the phase-space discontinuity shown by the vertical lines. 
	} 
	\label{fig:phase_particles}
\end{figure*}

\subsection{Particles}
\label{subsection:parts}

In this section we study the distribution of particles around clusters in $N$-body simulations. In Fig.~\ref{fig:phase_particles} we show the location of dark matter particles around clusters in the $r-v_{\rm rad}$ plane, where $r$ is the distance from the center of the cluster and $v_{\rm rad}$ is the radial velocity with respect to the center. The three top panels correspond to particle populations that were accreted onto the cluster at different times. The particles are separated based on when, in their orbital history, they first crossed within $4 ~{\rm  Mpc} ~h^{-1}$ of the host cluster\footnote{We choose $4~{\rm Mpc} ~h^{-1}$ to be safely outside the virial radius of the halo,  which is typically $\sim 1.5-2~{\rm Mpc}~h^{-1}$ for the halo masses considered in this work.}. The figure shows the stacked phase-space distribution of particles from cluster halos at $z=0$ and with $\langle M_{\rm 500c} \rangle = 3.1\times 10^{14} ~M_{\odot}h^{-1}$. This mass cut is chosen to match the mean mass of the  observed galaxy cluster sample described in Sec. \ref{sec:data_act}. Here we only focus on clusters at a single redshift for simplicity.

In the top, left panel we see particles that have been accreted onto the cluster halo most recently, with $z_{\rm acc}<0.25$, where $z_{\rm acc}$ is the redshift at which the particle crossed the boundary at $4$ Mpc $h^{-1}$. It appears that these particles mainly occupy the first ``infall'' stream in the region with $v_{\rm rad}<0$. This population corresponds to all particles that have infall times, $t_{\rm in}< t_{\rm max}$, where $t_{\rm max}$ is the lookback time corresponding to upper-limit on $z_{\rm acc}$, $3.1$ Gyrs. The infall or accretion time, $t_{\rm in}$, is also the time since crossing the $4 ~{\rm Mpc} ~h^{-1}$ boundary. The crossing time to the center of the halo from the boundary is $\sim 3 ~ \rm Gyrs$, indicating that these particles have mostly not crossed pericenter. On the other hand, if we increase the maximum time threshold, $t_{\rm max}$, to $5.6$ Gyrs (top row, middle panel),  such that we now have a population of particles that have been within the boundary for, \textit{at most}, 5.6 Gyrs, we find that they live in a stream that has wrapped around in phase space once. This population occupies the $v_{\rm rad}<0$ infall stream and also the region past pericenter with positive radial velocity, $v_{\rm rad}>0$. We note that these particles have not reached apocenter or splashback in their orbits (at which $v_{\rm rad}\sim 0 ~\rm km~s^{-1}$). If the maximum time threshold, $t_{\rm max}$, is increased further, to show the phase-space distribution of all particles that have been inside the cluster boundary for more than $9~\rm Gyrs$ for example, we find that particles have wrapped around in multiple streams (top, right panel), forming the complete virialized region of the halo.

The bottom row panels of Fig.~\ref{fig:phase_particles} show the spherically averaged 3D density, and its logarithmic slope as a function of radius, for the particles in the three panels of the top row. The slopes have been measured by smoothing the density profile using a Savitsky--Golay smoothing filter \citep{Diemer:2014xya}. We note, in the left panel, that the shape of the density profile changes as a function of the maximum infall time threshold. From the slope profiles, shown in the right panel, we infer that the density of particles in the infall stream (lightest grey curve), is well approximated by a single power law, but the same is not true for the other two cases. In fact, we find that the slopes show a distinct minimum in $r$ for particles that have been inside the halo longer than the time required for pericentric passage. The location of the minimum of the slope shifts for particles that have been accreted at different time thresholds. The white, dashed vertical lines in the top three panels show the location of the minimum evaluated from the slope profiles in phase space. In particular, we find that the minimum appears to trace the boundary between the multi-stream and single-stream region. This is most clearly demonstrated by comparing the middle and right panels in the top row; for the population of particles with $t_{in}<5.6 ~{\rm Gyrs}$, the white dashed line separates the two-stream region on its left at small $r$ from the single infall stream on its right, whereas for particles with $t_{in}<9 ~{\rm Gyrs}$, the minimum is at the location of the conventionally described splashback radius, i.e. at the boundary of the halo. The three particular values for the time thresholds were chosen to demonstrate the three distinct regimes. 

\textit{We conclude that the minimum of the slope traces a phase-space discontinuity that shifts for tracer populations accreted at different times. If we can separate populations of objects that have accreted onto halos at different times, we may expect to be able to measure this shift.}  Henceforth, we define a new generalized quantity, $r_{\rm smin}$, to refer to the location of the steepest slope.  This may be distinct from the traditional splashback radius for particle or galaxy populations that have not reached the apocenter of their orbits. We note that this result was discussed in the context of subhalos in \cite{Shin19}; here we verify that the trend holds for particle distributions as well. 

Here we use cumulative time bins or the maximum infall time thresholds, rather than differential bins, to connect with galaxy populations that enter the clusters at different times. For example, if we study the density of red galaxies around a cluster, we expect them to be present both in the infall stream and in the virialized region. In the next section we continue to study the distribution of the halos of such galaxies (subhalos in cluster halos) in simulations. 

\begin{table}
	\centering
	%\resizebox{}
	\begin{tabular}{ll}
		Variable & Definition \\ \hline
		$r_{\rm sp}$ & Location of the 3D splashback radius \\
        $r_{\rm smin}$ & Location of the minimum in the slope profile  \\
        $t_{\rm in}$ & time since infall/accretion onto the halo boundary\\
		$z_{\rm acc}$ & redshift at accretion \\
		$t_{\rm max}$ & maximum infall time threshold \\
		$t_{\rm d}$ & delay time in quenching model \\
		$t_{\rm q}$ & exponential quenching timescale in quenching model \\
	\end{tabular}
    \caption{Summary of important lengthscales and timescales used in the paper.}
\label{tab:variables}
\end{table}

\subsection{Subhalos}
\label{sec:sub_sims}
Dark matter halos form hierarchically, in the sense that small objects merge to form massive structures. A cluster halo is a relatively young object and contains a lot of existing substructure or subhalos that have not been destroyed through multiple orbits within its potential. The overall dynamics of halos in a cluster environment is also dictated by the cluster potential and is similar to that of the dark matter particles. All halos, if they are sufficiently massive, are expected to host galaxies at the centre of their potential wells, and therefore studying the evolution and distribution of substructure around clusters in simulations can help us understand the evolution of galaxies in observed clusters. As mentioned before, halos in our simulations are found using the code {\sc Rockstar} \citep{Rockstar}. The merger history of the halos are generated using the {\sc consistent-trees} algorithm \citep{Behroozi_2012}. Among other properties, the final halo catalog provides the redshift at the time of accretion, $z_{\rm acc},$ of a subhalo onto its parent host halo. This time is recorded when a subhalo crosses the virial radius of the parent cluster. In general we study the entire field of halos around each cluster; this includes objects that have been tagged as subhalos of the cluster in the halo catalog and also others in the infall region. Halos that have not crossed the virial boundary have $z_{\rm acc}=z_{\rm cluster}$.

We study the distribution of substructure/halos around clusters in the MDPL2 simulation. We select a sample of clusters that matches the distribution of mass and redshift of our observed SZ sample (\ref{sec:data_act}). All clusters lie in the redshift range $0.15 <z < 0.7$ and have a mean mass $\langle M_{\rm 500c} \rangle= 3.1\times 10^{14} M_{\odot}h^{-1}$.  We extract all halos that have peak maximum circular velocity $v_{\rm peak}>150$ km$~s^{-1}$ within a spherical volume of radius $5 ~{\rm Mpc} ~h^{-1}$ around the cluster centers at every redshift. The property $v_{\rm peak}$ is a proxy for halo mass, and is known to correlate best with the luminosity of the galaxy that it hosts \citep{Reddick:2012qy}.

\begin{figure}
    \includegraphics[width=0.9\linewidth]{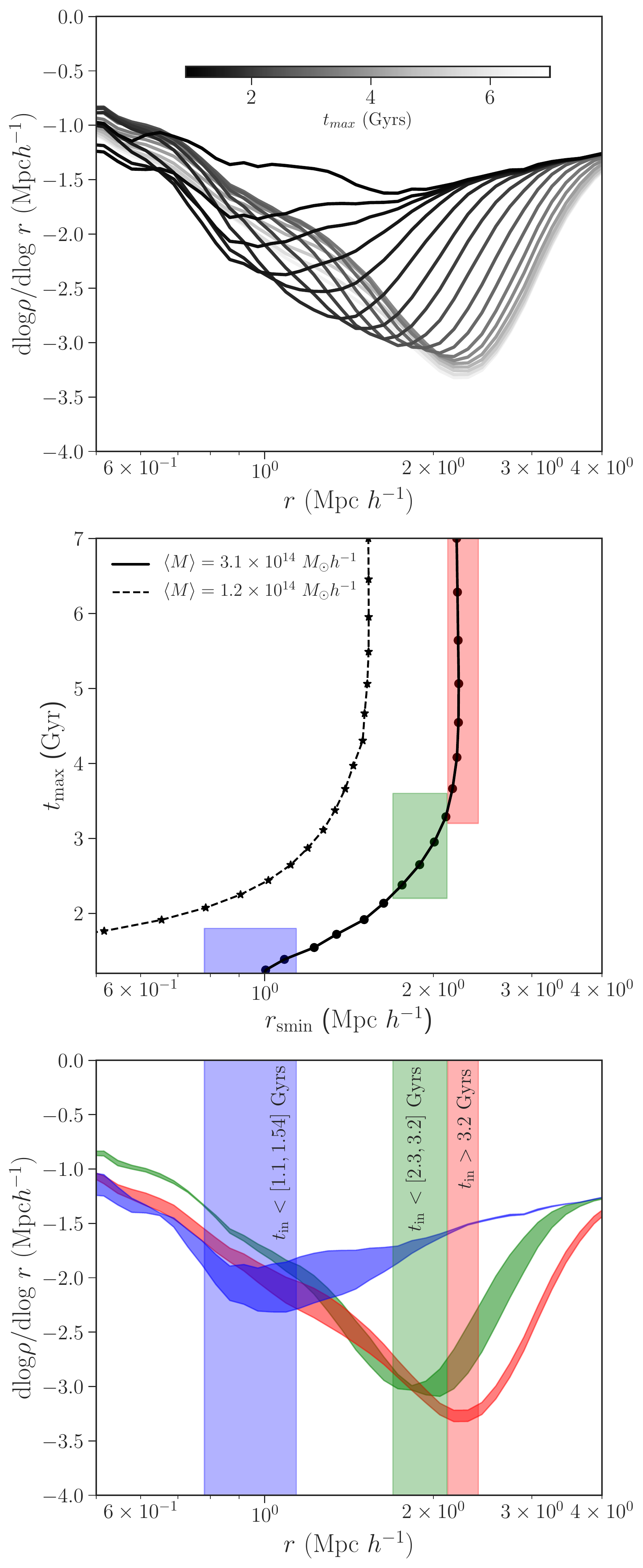}
    \caption{{\it (Top)} The slope of the density profile as a function of radius for populations of subhalos that have different maximum infall time thresholds. Each curve corresponds to the profiles of subhalos that have been inside the virial radius for at most $t_{\rm max}$ Gyrs, i.e. $t_{\rm in}<t_{\rm max}$ (objects that have not entered the host at all have $t_{in}=0$). {\it (Middle)} $t_{\rm max}$ of a population of subhalos as a function of the location of the minimum of the slope, $r_{\rm smin}$. The two separate curves correspond to different cluster mass samples. The width of the blue, green, and red bands correspond to the $1\sigma$ measurements of $r_{\rm smin}$ for observed blue, green, and red galaxies (Section \ref{sec:infall_time_obs}). {\it (Bottom)} Slope profiles for subhalo populations that have $r_{\rm smin}$ in the range of values obtained from data (shaded curves) compared to the estimated timescales shown in the middle panel.}
    \label{fig:splashback_subhalos}
\end{figure}

Halos that cross the virial radius of the cluster at different times occupy distinct regions in phase space in a manner similar to dark matter particles discussed in the previous section. We study the density profiles of these populations; in particular, we study how the 3D number density profile changes if we vary the maximum infall time, $t_{\rm max}$, of a given population of halos around a cluster. We compute the stacked 3D number density of halos in every case as a function of cluster-centric radius and measure the logarithmic slope of the profile using the Savitsky--Golay smoothing filter. We summarize our results in Fig. \ref{fig:splashback_subhalos}. In the top panel we show the logarithmic slope of the number density profile of a population of halos around clusters that have infall times, $t_{\rm in}<t_{\rm max}$;  each curve corresponds to a different $t_{\rm max}$. The radius $r_{\rm smin}$, is the location of the minimum in the slope-profile. The middle panel maps the movement of $r_{\rm smin}$ with $t_{\rm max}$.  Halo populations that contain objects that have been accreted earlier than $\sim 4\,{~\rm Gyrs}$ show a minimum in their profiles at the location of the cluster splashback radius, while $r_{\rm smin}$ falls off to smaller values for more recently accreted populations. This shows that the location of the slope-minimum, as is the case for particles, traces the location of discontinuity in phase space and contains information about the average infall time of a galaxy population. The dashed curve in the middle panel corresponds to the relation between $t_{\rm max}$ and $r_{\rm smin}$ for a sample of lower mass halos with $\left<M_{500c}\right>=1.2\times10^{14} ~M_{\odot} h^{-1}$.  The shape of the curve appears to be similar; the offset between the two curves arises from the overall smaller physical sizes of low mass objects, that scale as approximately as $M^{1/3}$.

 Dark matter particles are indistinguishable from each other, and probes of the matter distribution, like weak lensing, are sensitive to the total enclosed mass as opposed to populations of particles with different infall times. On the other hand, the observed properties of galaxies change with time after they are accreted onto clusters. Measuring the rate of change is often difficult because we do not know how long a particular population has been inside cluster. {\it The methods outlined above allow us to use the number density profile of galaxies and its slope to probe the infall history of galaxies, and provide a way to determine how long galaxies have been in a cluster.}  In the bottom panel of Fig. \ref{fig:splashback_subhalos}, we show the spread of allowed curves corresponding to the $r_{\rm smin}$ measured from  three different populations of galaxies with varying colors. The following discussion in the paper describes the connection between these two pictures. In the next section we use insights from the evolution of subhalos in simulations to understand galaxy evolution in observed galaxy clusters using ACT and DES data. 

\begin{figure}
	\includegraphics[width=0.95\linewidth]{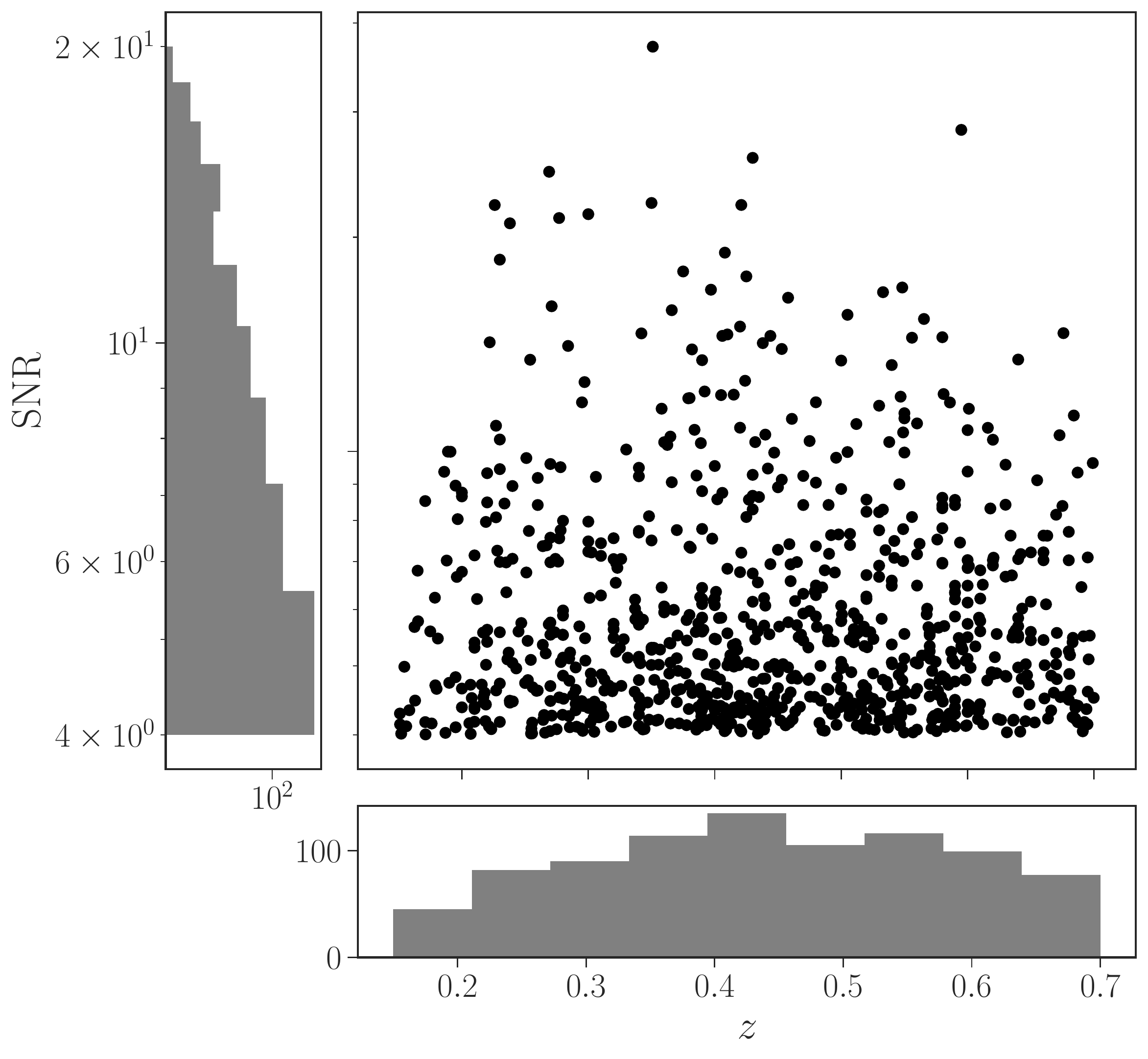}
	\caption{Distribution of ACT DR5 clusters in signal to noise (SNR) and redshift. Clusters with ${\rm SNR} > 4$ and $0.15 < z < 0.7$ within the DES-Y3 footprint, 908 in total, are shown in the figure and used in the analysis. The mean mass of the cluster sample is $3.1\times 10^{14} M_\odot h^{-1}$. }
	\label{fig:cluster_selection}
	\end{figure}

\section{Observations: Galaxy profiles and splashback radius}
\label{sec:obs}

In this section we describe the data and methods used for measuring the density distribution of galaxies and their splashback radius in massive galaxy clusters.  Our procedure
closely follows the method implemented in \citet{Shin19}; we refer readers to that work for details.  Here we briefly summarize the method to measure profiles, describe our color selection criteria, and present our results. 

\subsection{Data}

\subsubsection{{ACT cluster catalog}}
\label{sec:data_act}
We measure cluster--galaxy cross-correlations to estimate the splashback radius around massive galaxy clusters detected by the SZ effect from the Advanced Atacama Cosmology Telescope (AdvACT) survey. The cluster catalog is part of the fifth data release from ACT \citep[ACT DR5, ][]{ACTcluster_inprep}. The cluster catalog is derived from applying a multi-frequency matched filter \citep[e.g.][]{Melin06,Williamson11} to 98 and 150\,GHz ACT day- and night-time observations taken from 2008--2018. For this work, the ACT maps for each observing season and detector array were co-added using the procedure described in \citet{Naess:2020wgi}. 

The cluster signal is modeled using the Universal Pressure Profile \citep[UPP;][]{Arnaud10}, and masses are inferred from the SZ signal in a method similar to \citet{Hilton18}. Here we make use of masses that have been rescaled according to a richness-based weak-lensing mass calibration procedure by a factor of $0.69\pm 0.07$ \citep[see][]{Hilton18}. The survey selection function is defined using maps filtered at a single reference filter scale ($2.4^{\prime}$, equivalent to a cluster with $M_{\rm 500c} = 2 \times 10^{14}$\,M$_{\odot}$ at $z = 0.4$). We apply a cut on signal-to-noise ratio, $SNR>4$, and  redshift, $0.15 < z < 0.7$, which gives a total number of 908 clusters in the $4552~{\rm deg}^2$ overlapping area between DES and ACT. The mean cluster mass for this sample is  $3.1 \times 10^{14} \, h^{-1}M_{\odot}$.
The redshift and SNR distribution of the SZ clusters is shown in Fig.~\ref{fig:cluster_selection}. The redshifts for the clusters have been obtained by compiling spectroscopic redshifts, where available, and preexisting catalogs that have overlap with ACT (see \citealt{ACTcluster_inprep} for a detailed description).

We also require a mock cluster catalog that closely follows the survey geometry of ACT, located at random positions in the sky to reliably measure cluster--galaxy cross-correlation functions. The mock ``cluster randoms'' were generated from the SZ-signal noise map generated by the cluster finder, by sampling from the \citet{Tinker08} mass function, and applying an SZ-signal--mass relation adjusted to reproduce the observed number of SZ signal to noise greater than 6 clusters found in the DES footprint \citep{ACTcluster_inprep}. 

\subsubsection{DES galaxy catalog}
\label{sec:data_des}
To measure the galaxy distribution around the selected clusters we use galaxies observed in DES. DES is a $5000 ~{\rm deg}^2$ survey that covers the Southern Galactic cap. The survey used the 570 megapixel Dark Energy Camera \citep{Flaugher:2015pxc}, mounted on the 4m Blanco Telescope in Chile, to imaged the sky in $grizY$ filters. In this work we use data from the first three years of observation, in particular the DES Y3 gold catalog \citep{DESY3GOLD_inprep} similarly generated as in \citet{DrlicaWagner17}. The image-processing pipeline used in DES is described in \citet{Morganson:2018pdr}. We use the galaxy magnitudes that have been corrected for differential reddening across the DES footprint. After excluding galaxies with extreme colors (outside the range $-1<(g-r)<3$, $-1<(r-i)<2.5$, $-1<(i-z)<2$), we take all galaxies with $i$-band magnitude, $m_{\rm i}$, smaller than 22.5 and only use parts of the footprint for which the depth of the survey in $m_{\rm i}$ is larger than 22.5 to ensure the completeness; this leaves $\sim$90\% ($\sim 4500 ~\rm deg^2$) of the entire DES footprint. We also require all galaxies to have uncertainties on the magnitude smaller than $0.1$. When calculating the cross-correlation function between the clusters and the galaxies, we apply a further limit on the absolute magnitude, $M_{\rm i}<-19.87$, which corresponds to the apparent magnitude limit of $m_i<22.5$ at the maximum redshift of $z=0.7$. This is to ensure the same luminosity cut on the galaxies regardless of the redshift of the clusters. The photometric redshifts for the galaxies are estimated with Directional Neighbourhood Fitting (DNF) algorithm \citep{DNf}. The galaxy sample used in this paper is almost identical to the one used in \citet{Shin19}; we refer the readers there for further details.

\subsection{Method}
\subsubsection{Color selection in data}
\label{sec:color_split}

\begin{figure}
    \includegraphics[width=1\linewidth]{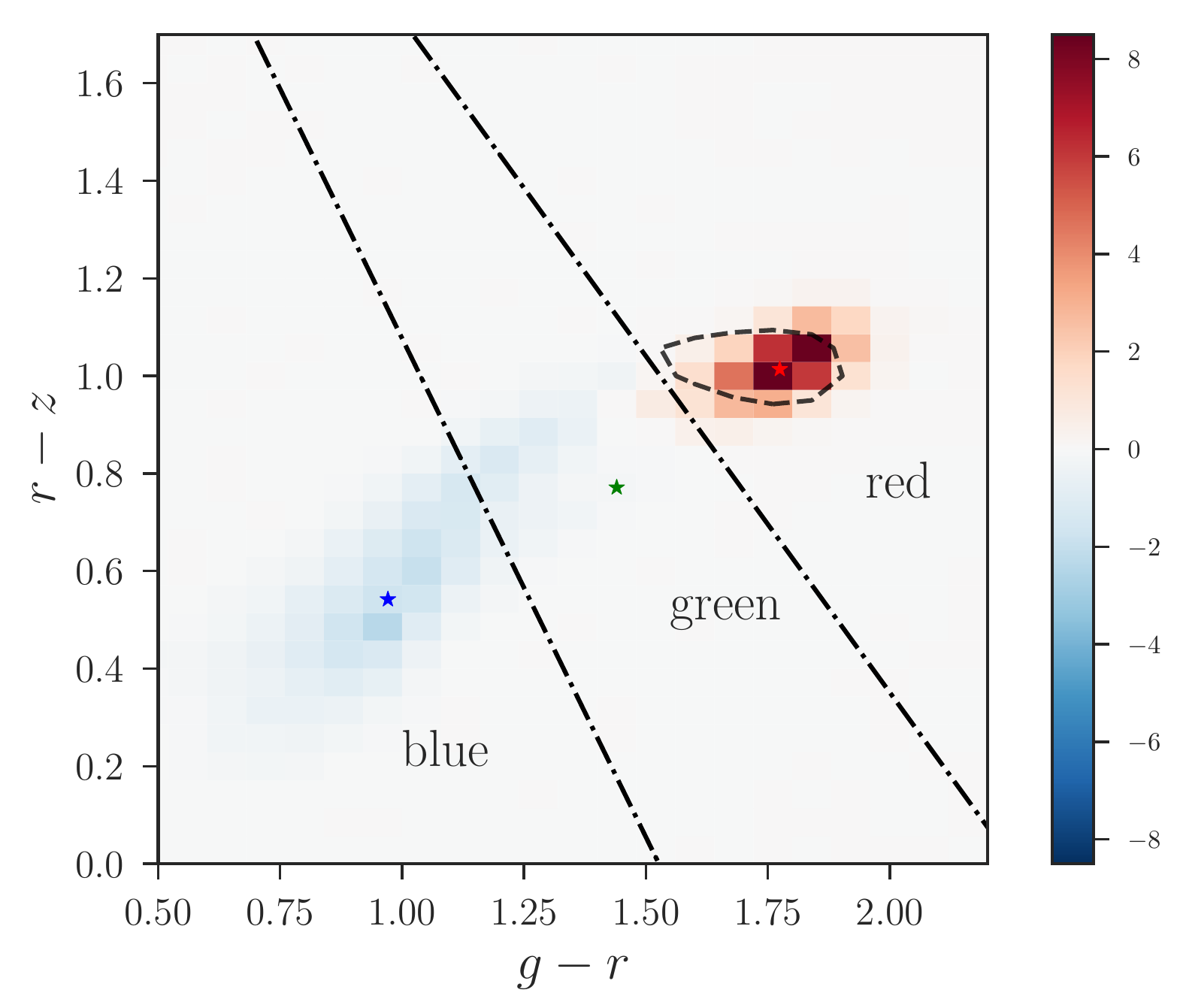}
    \caption{Color-selection method in DES data. The plot shows the excess number density of galaxies in the color--color plane around 1~${\rm Mpc}~ h^{-1}$ of the ACT DR5 clusters (with respect to random locations on the sky). The dot--dashed lines are used to assign galaxy colors, red/green/blue, from upper right to lower left, respectively. See Sec.~\ref{sec:color_split} for details. For comparison, the black dashed contour marks the 68$\%$ range of the red-sequence galaxies drawn from the optically selected clusters (RedMaPPer) in DES data.}
    \label{fig:color_selection}
\end{figure}

Fig. \ref{fig:color_selection} shows how DES galaxies are assigned to different color bins. 
We measure the density of galaxies in the $(g-r)$--$(r-z)$ color space in redshift bins of $\Delta z = 0.05$. 
Specifically, we measure the density of galaxies in color space around our cluster sample, within $1 ~{\rm Mpc}~h^{-1}$ of the center, and also around random points on the sky; we then subtract the latter from the former. 
The plot shows the resultant map of ``overdensity'' in the color space around our cluster sample, with respect to the global density, in the redshift bin of $z=[0.45,0.50]$ as an example.
One can see an excess of red galaxies and a deficit of blue galaxies around the clusters, due to quenching of galaxy star formation inside the clusters.
This tendency is prevalent in every redshift bin.

We define color bins for the galaxies as follows. 
First we identify the ``red peak'' as the average overdensity-weighted location of the five points with the largest values of overdensity (red point). 
We identify the ``blue peak'' similarly, with the five points of the smallest overdensity values (blue point). 
We finally define the green valley as the location where the absolute value of overdensity is minimized between the red peak and the blue peak (green point).
We then draw two lines that pass through (1) the midpoint between the red peak \& the green valley and (2) the midpoint between the green valley \& the blue peak.
These lines are also perpendicular to the line adjoining the red peak \& the green valley and the line adjoining the green valley \& the blue peak, although the specific choice of the slope of the lines does not affect our color selection significantly since most of the cluster member galaxies are located around the narrow path in the color--color space, as shown in the figure.
These two lines then separate the red/green/blue galaxies (three regions divided by the two lines).
The black contour in the figure includes the 68\% of the red-sequence galaxies drawn from the member catalog of the optically selected (\texttt{RedMaPPer}, \citealt{Rykoff2014}) clusters in the DES, which identifies the clusters by detecting red-sequence overdensities at each redshift.
The agreement between the locations of the red-sequence galaxies and the calculated red peak is evident.

\subsubsection{Modeling galaxy number density profiles}

We model the  profiles of galaxies using the fitting formula for galaxy number density profiles described in  \cite{Diemer:2014xya} and used in subsequent work on splashback in different cluster and galaxy samples \citep{More:2016vgs, Baxter:2017csy, Chang:2017hjt, Shin19, Zuercher:2018prq}. The 3D number density profile is defined by an inner, virialized Einasto component \citep{Einasto1965} and an outer, infall, two-halo like term along with a transition region\footnote{see Section 3.3 in \citet{Diemer:2014xya}}.

\begin{equation}
\rho(r)= \rho_{\rm inner}(r)f_{\rm trans}(r)+\rho_{\rm out}(r)    
\end{equation}

where,

\begin{equation}
\rho_{\rm inner}(r)= \rho_{\rm s} \exp\left(-\frac{2}{\alpha} 
\left[\left(\frac{r}{r_{\rm s}}\right)^\alpha-1\right]\right)
\end{equation}

\begin{equation}
    f_{\rm trans}(r)=\left[1+\left(\frac{r}{r_{\rm t}}\right)^\beta \right]^{-\gamma/\beta}
\end{equation}

\begin{equation}
    \rho_{\rm outer}(r)=\rho_0\left(\frac{r}{r_0}\right)^{-s_{\rm e}}
\end{equation}
The 3D density is integrated to obtain the expected projected density, $\Sigma (R)$, as a function of projected radius, $R$, with the maximum projection length of $40 ~{\rm Mpc}~ h^{-1}$. 

We also account for the cluster miscentering. Due to the finite beam size in the CMB survey and other systematic effects, the calculated center of the clusters could differ from the true cluster center of mass. 

The azimuthally averaged profile of a cluster miscentered by a distance $R_{\rm mis}$, is		
\begin{equation}
\Sigma_{\rm mis}(R|R_{\rm mis}) = \\
\int^{2\pi}_0 \frac{{\rm d}\theta}{2\pi}\Sigma_0\Big(\sqrt{R^2 + R^2_{\rm mis} + 2RR_{\rm mis}{\rm cos}\theta}\Big),
\end{equation}
where $\Sigma_0$ is the profile without miscentering.
We average the profiles over the distribution in $R_{\rm mis}$ as
\begin{equation}
\Sigma_{\rm mis} (R) = \int {\rm d}R_{\rm mis}P(R_{\rm mis})\Sigma_{\rm mis}(R|R_{\rm mis}),
\end{equation}
where $P(R_{\rm mis})$ is the probability distribution of a cluster to be miscentered by a distance $R_{\rm mis}$ from the true center, which we model as a Rayleigh distribution \citep{Saro15}:
\begin{equation}
\label{eq:miscenend}
P(R_{\rm mis}) = \frac{R_{\rm mis}}{\sigma^2_R} {\rm exp}\left[-\frac{R^2_{\rm mis}}{2\sigma^2_R}\right],
\end{equation}
where $\sigma_R$ characterizes the width of the $R_{\rm mis}$ distribution. Miscentering is generally measured as a fraction of the FWHM of the ACT beam, which has a fixed angular size; therefore the radius of miscentering is a function of the cluster redshift. In this paper we treat $\sigma_{\rm R}$ as a free parameter initialized at $\sigma_{\rm R}=0.13 ~{\rm Mpc}~h^{-1}$ in radial units, with a wide prior range. We have verified that our choice of prior is a conservative one and is consistent with an angular width of $0.3$ arcmin for $SNR=4$ clusters at the mean redshift of the sample \citep{ACTcluster_inprep}.

Our model has a total of nine parameters, eight from the halo model and one from the miscentering model. The parameters $r_0$ and $\rho_0$ are degenerate with each other, therefore we fix $r_0=1.5$ Mpc $h^{-1}$. We fit the projected surface number density profile, described in Sec. \ref{sec:measurement}, with the nine-parameter model above using a Markov Chain Monte Carlo (MCMC) method implemented in \texttt{emcee} package \citep{emcee13}, with priors described in the Table~\ref{tab:modeling_parameters}. In comparison to the priors assumed in \citet{Shin19}, we use 

(1) the same prior on the Einasto slope parameter $\alpha$ since the mean mass is the same ($3.1 \times 10^{14} \, h^{-1}M_{\odot}$), (2) three times wider prior ranges on the slope parameters of $f_{\rm trans}(r)$, ($\beta$, $\gamma$) to allow more flexibility on the fitting especially for the blue and green galaxies and (3) three times wider prior on the miscentering parameters.

\begin{table}
	\centering
	\begin{tabular}{ll}
		Parameter & Prior \\ \hline
		$\log \rho_{\rm s}$ & $[-\infty,\infty]$ \\
        $\log \alpha$ & $\mathcal{N}(\log(0.22),0.6^2)$ \\
        $\log r_{\rm s}$ & $[\log(0.01),\log(5.0)]$ \\
		$\log r_{\rm t}$ & $[\log(0.1),\log(5.0)]$ \\
		$\log \beta$ & $\mathcal{N}(\log(6.0),0.6^2)$ \\
		$\log \gamma$ & $\mathcal{N}(\log(4.0),0.6^2)$ \\
		$\log \rho_{\rm 0}$ & $[-\infty,\infty]$ \\
        $s_{\rm e}$ & [0.1,10.0] \\
        $\ln \sigma_{\rm R}$ & $\mathcal{N}(-2.0,1.2^2)$ \\
	\end{tabular}
    \caption{Prior range of each model parameter. $\mathcal{N}(m,\sigma^2)$ represents a Gaussian prior with  mean $m$ and  standard deviation $\sigma$. The densities, $\rho_{\rm s}$ and $\rho_{\rm t}$ are in units of Mpc$^3$ $h^{-1}$ and radii $r_{\rm s}$, $r_{\rm 0}$ are in units ${\rm Mpc}~h^{-3}$. The miscentering parameter $\sigma_{\rm R}$ is also in units of Mpc $h^{-1}$.}
\label{tab:modeling_parameters}
\end{table}

\subsubsection{Measurement and  MCMC fitting of galaxy profiles}
\label{sec:measurement}

We adopt the same method to measure the galaxy surface number density profile, $\Sigma_{\rm g}$, as implemented in \citet{Shin19}. We briefly summarize the procedure here and refer readers to \citet{Shin19} for details.

The galaxy surface density profile, $\Sigma_{\rm g}$, around the clusters is related to the cluster--galaxy cross-correlation function, $\omega(R)$, as
\begin{equation}
    \Sigma_{\rm g} = \bar{\Sigma}_{\rm g} \omega(R),
\end{equation}
where $\bar{\Sigma}_{\rm g}$ is the mean galaxy number density.
We first divide the clusters into redshift bins of width $dz=0.025$. In each redshift bin, we measure the cluster--galaxy cross-correlation function, $\omega(\theta,z_i)$, using the Landy--Szalay estimator \citep{Landy93}. The angular cross-correlation function, $\omega(\theta,z_i)$, is then converted to $\omega(R,z_i)$ assuming the midpoint redshift value of the bin. We finally average $\omega(R,z_i)$ over the redshift bins weighted by the number of clusters in each bin to obtain $\omega(R)$. Then we multiply it with the mean galaxy density to generate the final estimate of $\Sigma_{\rm g}$. We use fifteen bins between $0.1 < R < 20 ~{\rm Mpc}~h^{-1}$ spaced equally on a logarithmic scale. 

The covariance matrix of the galaxy surface density profile is derived by the jackknife resampling method \citep{Norberg09} with 100 patches of similar size. Each jackknife patch retains $\sim$4.4$^2$ square degrees of area. The length of each patch corresponds to $\sim$100 ${\rm Mpc}~h^{-1} $ at $z=0.5$, significantly larger than the maximum distance scale of interest in this paper.

With the covariance matrix estimated by jackknife method, $\textbf{C}$, we assume the likelihood ($\mathcal{L}$) to be Gaussian. Given the data, $\vec{d}$, and the model parameters, $\vec{\theta}$ (Table \ref{tab:modeling_parameters}), the likelihood is written as:
\begin{equation}
\ln \mathcal{L}[\vec{d} | \vec{m}(\vec{\theta}) ] = -\frac{1}{2}\left[\vec{d} - \vec{m}(\vec{\theta})\right]^{\rm T} \textbf{C}^{-1} \left[\vec{d} - \vec{m}(\vec{\theta}) \right],
\end{equation} 
where $\vec{m}(\vec{\theta})$ is the model evaluated at the parameter $\vec{\theta}$. The posterior on the model parameters is then expressed as 
\begin{equation}
\ln \mathcal{P}(\vec{\theta} | \vec{d}) = \ln \big[ \mathcal{L}(\vec{d} | \vec{m}(\vec{\theta}) ){\rm Pr}(\vec{\theta}) \big],
\end{equation}
where ${\rm Pr}(\vec{\theta})$ are the priors imposed on $\vec{\theta}$.

\subsection{Results}
\subsubsection{Splashback radius in clusters from ACT DR5}

\begin{figure}
    \includegraphics[width=1\linewidth]{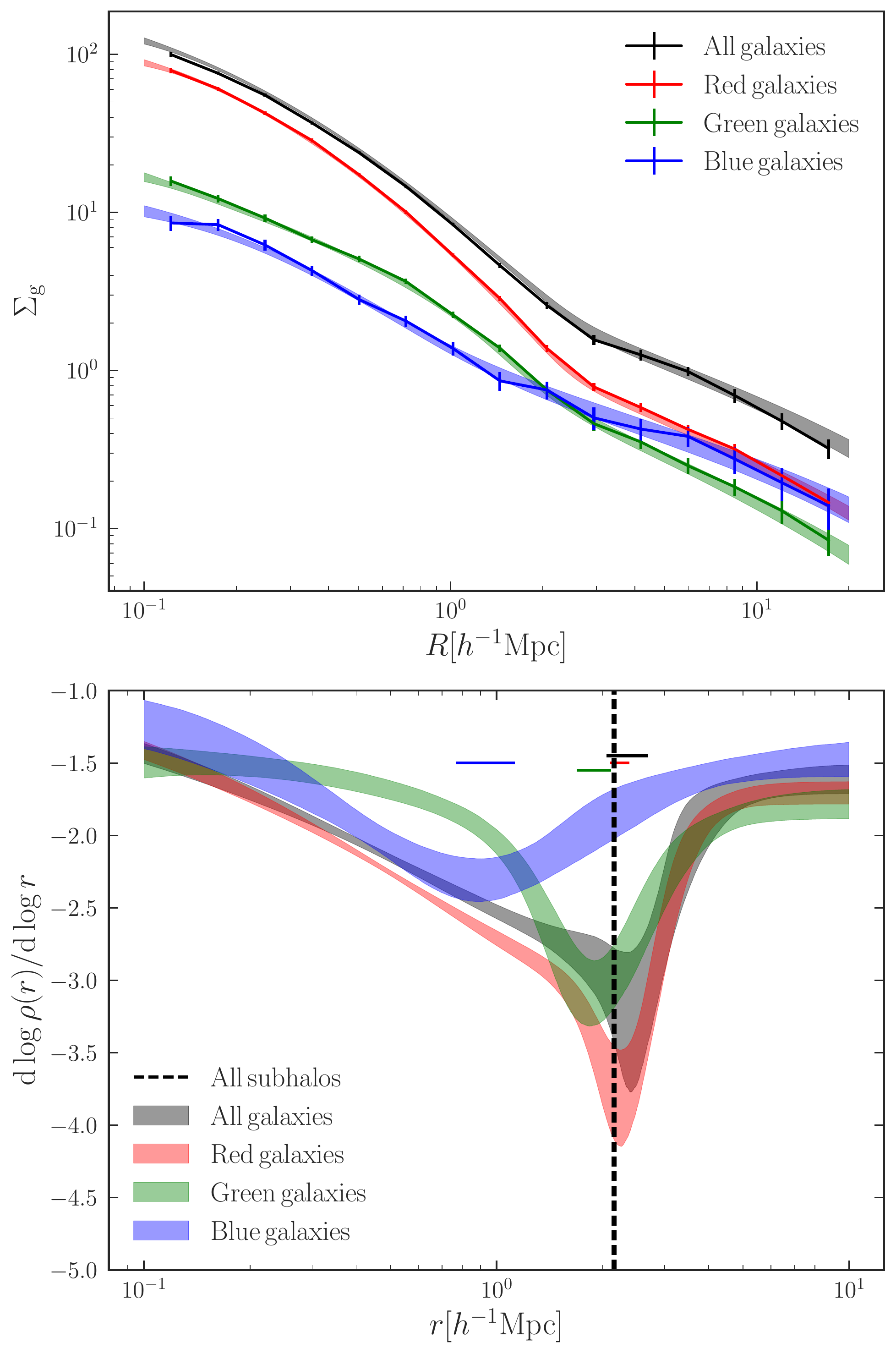}
    \caption{{\it (Top)} Measured surface density profiles of  galaxies around clusters from ACT DR5 as a function of 2D projected radius.  The shaded regions in red, green, and blue show the 1$\sigma$ ranges of the fitted profiles of galaxies of the corresponding color from the MCMC runs (see Sec.~\ref{sec:measurement}). {\it(Bottom)} Fitted 3D slope of the density profile as a function of galaxy color. The vertical line specifies the location of the splashback radius measured from  subhalos in the simulation (see Sec.~\ref{sec:splashback_measurement}). The colored segments (around the vertical line) show the estimated splashback radius from the data and its 1$\sigma$ uncertainty.}
    \label{fig:color_splashback_data}
\end{figure}

The top panel of Fig. \ref{fig:color_splashback_data} shows the measurement of the projected galaxy number density as a function of radius. Measurement errors are shown as solid curves. The shaded regions show the model fit to data obtained using the method described in the previous section; this region encompasses all curves within $1\sigma$ of the best-fit model parameters. The bottom panel shows the logarithmic slope of the corresponding 3D density profile obtained from the fitted parameters. The errors here are also derived from the $1\sigma$ values of the model parameters.

The black curve and the grey shaded region in the top panel corresponds to the projected number density and the model-fit for ``all'' galaxies around the cluster. The grey shaded region in the bottom panel shows the corresponding, fitted, 3D slope profile. The steepest slope measured from the profile is located at the radius $r_{\rm smin}=2.4^{+0.3}_{-0.4}$ Mpc $h^{-1}$. The black vertical line denotes the theoretical prediction for the location of the steepest slope measured from the density profile of halos with $v_{\rm peak}>150 ~\rm km~s^{-1}$ around matched clusters in the MDPL2 simulation. The $v_{\rm peak}$ threshold has been derived by matching the magnitude of the number density profile at large radius \citep{More:2016vgs}. 

When all halos are considered without splitting on accretion time, the transition from the infall region to the multistreaming region occurs at the traditional splashback radius. Therefore, the location of the steepest slope measured using all galaxies around the clusters corresponds to the splashback radius, $r_{\rm sp}$, for the sample. We find that the observed value is statistically consistent with theoretical predictions from simulations. We also confirm, in agreement with previous results from \citet{Shin19} and \citet{Zuercher:2018prq}, that the splashback radius measured in SZ clusters does not show any significant discrepancy between theory and observations (see black line and grey shaded curve in Fig. \ref{fig:color_splashback_data}). This is unlike the splashback radius measured using optically selected RedMaPPer clusters \citep{More:2016vgs, Baxter:2017csy, Chang:2017hjt, Shin19}, where the measured value is smaller than that expected from theory. A summary of the measurements of the splashback radius from different cluster samples with a range of masses is shown in Table \ref{tab:rsp}.

\begin{table}
	%\centering
	\resizebox{1.\columnwidth}{!}{
	\begin{tabular}{cccc}
		Sample & $[\left<M_{200m}\right>, \left<z\right>]$ & $r_{\rm sp}$ (Mpc $h^{-1}$) & Reference \\ \hline
        SDSS RM & [$1.9e14$, 0.24] & $1.18\pm0.08$  & \shortstack{\citet{More:2016vgs} \\ \citet{Baxter:2017csy}}  \\
        DES RM & [$1.8e14$, 0.41] & $1.13\pm0.07$ & \citet{Chang:2017hjt} \\ 
        DES RM$^*$ & [$1.8e14$, 0.41] & $1.4\pm0.2$ & \citet{Chang:2017hjt}\\ 
     	SPT SZ & [$5.3e14$, 0.49] & $2.4^{+0.5}_{-0.5}$ & \citet{Shin19} \\
        ACT SZ & [5.8e14, 0.49] & $2.2^{+0.7}_{-0.6}$  & \citet{Shin19}\\
        Planck SZ & [6.2e14, 0.177] & $1.9^{+0.4}_{-0.3}$  & \citet{Zuercher:2018prq}\\
        ACT DR5 SZ & [5.8e14, 0.49] & $2.4^{+0.3}_{-0.4}$ & This work\\
 
	\end{tabular}
	}
    \caption{Comparison of the measurement of the location of splashback radius from all galaxies with earlier work. RM corresponds to the RedMaPPer cluster sample and SZ to Sunyaev--Zeldovic samples. Each row designates the splashback radius measured using the galaxy number density profiles except row 3, DES RM$^*$, which specifies the splashback measured using weak lensing. The mass is quoted in units of $M_\odot h^{-1}$.}
    \label{tab:rsp}
\end{table}

\subsubsection{Galaxy density and slope profile as a function of galaxy color}
\label{sec:splashback_measurement}

In this section we study the distribution of galaxies of different colors around the clusters from the ACT DR5 catalog. The red, green, and blue curves in Fig. \ref{fig:color_splashback_data} show the measurements and their best-fit model curves for galaxies assigned to the corresponding color bins using the method described in section \ref{sec:color_split}.

We find that the overall shape of the projected number density profiles, and consequently the slope profiles, is different for the three populations. The apparent location of the splashback radius, traditionally defined as the location of the minimum of the slope profile, appears to show shifts as a function of galaxy color. 
The locations of the slope minimum, $r_{\rm smin}$, for the red, green, and blue galaxies are constrained to $2.2^{+0.2}_{-0.1}$, $1.9^{+0.2}_{-0.2}$ and $0.95^{+0.2}_{-0.1}$ Mpc $h^{-1}$, respectively.
The slope-minimum for green galaxies is shallower and at a smaller radius compared to red ones, but the difference is not statistically significant; however the blue galaxies show a weak feature, with the minimum in the slope located at a smaller radius compared to both green and red galaxies. We also note that the red galaxies show a significantly steeper inner profile compared to green and blue galaxies. 

Previous work has also measured the slope of the density profile as a function of galaxy color. \citet{Baxter:2017csy} measured the splashback radius in the bluest and reddest quartiles of $(g-r)$ galaxy colors for RedMaPPer clusters and found that the splashback feature was more prominent for galaxies in the red quartile and slightly larger than the full galaxy sample, while the galaxies in the blue quartile show a weak splashback-like feature. \citet{Shin19} developed an improved color selection and measured the splashback locations of galaxies of different colors around SZ-selected clusters from the South Pole Telescope (SPT) and ACT.
\citet{Shin19} looked at not only galaxies in the blue and red quartiles, but also galaxies that lie in the green valley. While the bluest galaxies did not show a significant minimum in the transition region, the green and red galaxies both show a distinct splashback-like minimum feature. Due to the small number of clusters (300 for the SPT sample and about 100 for the ACT sample) small shifts of the minimum between galaxies of different colors, if they exist, could not be detected. In the current sample of ACT DR5 clusters, with three times the number of clusters compared to the SPT sample, we can study more subtle changes in the density profiles with color. \\

Considering that galaxies evolve in clusters due to astrophysical processes, with their SFRs decreasing with time, we expect galaxies to evolve from being blue and star-forming to red and quenched over the course of their orbits, migrating between the different color bins. The density profiles that we measure here are a snapshot in time of this process. In Sec.~\ref{sec:splashback_sims} we studied the signatures of the net time spent within a cluster on the density profile of a population of halos. In particular we found that the location of the minimum of the slope of the density profile traces a phase-space discontinuity and encodes information about how long a population of subhalos has been inside a halo. Therefore, one possible explanation for the movement of the location of the steepest slope for galaxies of different colors observed in data can be the movement of the phase-space boundary for the different populations as they evolve through the phase space of the halo. In the next section, we elaborate on the possible connection between galaxy colors and infall time and attempt to model the process of galaxy quenching within clusters by mapping galaxies to subhalos in $N$-body simulations.

\section{Galaxy quenching in clusters}
\label{sec:quenching}

\subsection{Infall time of observed galaxies}
\label{sec:infall_time_obs}

If intra-cluster processes are indeed responsible for quenching the SFRs of galaxies that fall into clusters, we may expect that an infalling population of galaxies will evolve to a redder population over time. Considering that quenching is a continuous process,  galaxies that are originally blue will convert to green galaxies and eventually to red ones, while green galaxies will turn red with time.\footnote{We reiterate that by green galaxies we mean galaxies that are in the green valley, between red galaxies with low SFRs and blue ones with high SFRs} This implies, firstly, that the number of galaxies as a function of color is not conserved at each point in the phase space of clusters. Secondly, if the quenching timescales are short enough, blue galaxies will not exist in the phase-space locations populated by galaxies that have been within the cluster over multiple orbits, therefore affecting the density profiles in ways discussed in Sec. \ref{sec:splashback_sims}. For example, if all blue galaxies are quenched before pericentric passage or at pericenter, they will only exist in the single infall stream and won't show a minimum in their logarithmic slopes. Similarly if all green galaxies convert to red ones before they reach the apocenter of their first orbits (i.e. splashback), $r_{\rm smin}$ will be smaller compared $r_{\rm sp}$. In this scenario, the galaxies that reach splashback are only red and the steepest slope in their profile $r_{\rm smin,red}\equiv r_{\rm sp}$.

We compare $r_{\rm smin}$ observed for different color bins in the observed DES galaxy population with $r_{\rm smin}$ for halo populations in simulations that have been accreted onto their host clusters at different times on average. These comparisons are shown in Fig. \ref{fig:splashback_subhalos}. The middle panel of the figure demonstrates that different values of $r_{\rm smin}$ correspond to populations of halos that have a different maximum infall time threshold, $t_{\rm max}$. Each $r_{\rm smin}$ is derived from the logarithmic slope of the number density profile of halos by varying the quantity $t_{\rm max}$, such that each population comprises all halos around clusters with $t_{\rm in}<t_{\rm max}$ (top panel). The horizontal width of red, green, and blue shaded regions in the middle panel of Fig. \ref{fig:splashback_subhalos} covers the $1\sigma$ region  of $r_{\rm smin}$ for the galaxies with different colors measured from data. Comparing with the simulations, we find that red galaxies in data show an $r_{\rm smin}$ consistent with halos in the simulation that have been inside the halo for longer than $3.2$ Gyrs. The $r_{\rm smin}$ for green galaxies corresponds to the $r_{\rm smin}$ for halos in simulations that have been in the cluster for at least $t_{\rm in} > 2.2$ Gyrs; they most likely do not survive beyond $3.2$ Gyrs. The blue galaxies are consistent with a population of objects that have been inside the cluster for less than $1.5$ Gyrs. The bottom panel of the figure shows the simulation curves for the slope profiles for halo populations that have their $r_{\rm smin}$ in the $1\sigma$ region of $r_{\rm smin}$ for observed galaxies.

We conclude that the shift in the minimum of the slope of the density profile for observed galaxies, can be inferred to arise from the phenomenon that different populations of galaxies survive in the halo for different net amounts of time. \textit{We emphasize that this is not in fact a movement of the ``splashback radius,'' or the boundary of the halo, traced by the apocenter of the first orbit for different galaxy colors but instead a movement of the phase-space boundary for populations of different colors}. In other words while the ``splashback'' radius traces the boundary of the whole multi-streaming region of the halo, different color galaxies, by virtue of the fact that galaxies change color over their orbits, can have their streams end at different locations in phase space. 

\subsection{Modeling galaxy quenching}
\label{sec:model_quench}
In this section we model intra-cluster galaxy quenching to obtain the relevant timescales involved in star-formation quenching within clusters. We adopt the commonly used exponential quenching model \citep{Papovich:2001bu, Shapley:2005er, Lee:2010we, Schreiber:2009fi}; the details of our model follow  \citet{Wetzel:2012nn}. In this model the quenching of star formation in a galaxy falling into a cluster is described by two relevant timescales, a delay-time, $t_{\rm d}$, which is the duration of time that the galaxy SFR remains unaffected by intra-cluster processes, and an exponential decay timescale, $t_{\rm q}$, after the period of delay has passed. The outskirts of clusters have low densities of both gas and dark matter, it can be conjectured  therefore that infalling galaxies start quenching strongly only when they reach the central regions of the host; the delay time, $t_{\rm d}$, can therefore be related to the pericenter crossing time of the galaxy. While $t_{\rm d}$ and $t_{\rm q}$ capture star-formation quenching \textit{within} the cluster, we note that galaxies also have an intrinsic decay time for their SFR in the field \citep{Noeske:2007hr} , i.e. isolated  or central galaxies (galaxies in the field) eventually stop forming stars with time. We call this intrinsic decay timescale, $t_{\rm q,\rm iso}$, the quenching timescale for isolated (field) galaxies;
we assume that this timescale applies to all galaxies before they are impacted by cluster processes.

Adopting the model from \citet{Wetzel:2012nn} , the SFR of a satellite galaxy that falls into the cluster can be defined as, 

\begin{equation}
    SFR_{\rm sat}(t) =\begin{cases} SFR_{{\rm iso}}(t), & \mbox{if}~t<t_{\rm d} \\ SFR_{\rm iso}(t)~{\rm exp}\left(-\frac{t-t_{\rm d}}{t_{\rm q}}\right) &\mbox{if} ~t> t_{\rm d}, \end{cases}
\label{eqn:quenching_model}
\end{equation}
where $SFR_{\rm sat}(t)$ represents the SFR of galaxies that are satellites of clusters at time $t$, $SFR_{\rm iso}(t)$ is the SFR of isolated galaxies, $t$ is the time since infall of the galaxy into the cluster virial radius. 

The intrinsic evolution of SFR of an isolated galaxy is modelled as 

\begin{equation}
    SFR_{\rm iso}(t) = SFR(t_{\rm i})~{\rm exp}\left(-t/t_{\rm q,\rm iso}\right),
\label{eqn:field_evn}
\end{equation}
where, $t_{\rm q, iso}$ is the intrinsic timescale for the decay of star formation, independent of the cluster environment, and $SFR(t_i)$ the initial SFR when the galaxies are formed at time $t_i$. 

We obtain constraints on our model for quenching in two separate ways. Firstly, we use only the information about the location of the splashback radius and secondly, we use the profile of the ratio of galaxy number densities of different colors. In the following sections we describe our method.

\subsection{Constraints on galaxy quenching timescales}

Our goal is to constrain the galaxy star-formation evolution timescales, $t_{\rm d}$ and $t_{\rm q}$, in our model. The time $t_{\rm d}$ is the delay before quenching begins inside a cluster, and $t_{\rm q}$ is the exponential decay rate of star formation after the onset of quenching.  To constrain the quenching timescales we make mock galaxy catalogs from the MDPL2 simulations by assigning galaxies to halos around cluster-mass objects, and evolve every mock galaxy in our sample with the model described above to find the parameters that best describe our data. We match the redshift distribution and mean mass of the simulation cluster sample to the observed sample of SZ clusters. We reiterate that we refer to the main host cluster dark matter halo as a ``cluster'' and study the entire halo field around it, including both subhalos and halos outside the virial radius of the cluster. 

Firstly, we select a minimum $v_{\rm peak}$ threshold for the halos in the simulation to correspond to the magnitude limit of the observed galaxies. We use abundance matching to determine this threshold, i.e. the cumulative distribution of observed galaxy magnitudes is matched to the $v_{\rm peak}$ distribution of halos in simulation.\footnote{We use peak quantities for subhalos under the assumption that the galaxy stellar mass or luminosity traces the mass of the original unstripped subhalo before it falls into the cluster's tidal field.} We find that our sample of galaxies with $M_r < -19.87$ matches with a halo $v_{\rm peak}$ threshold of $150 ~\rm km~s^{-1}$. We extract all subhalos with $v_{\rm peak}>150 ~\rm km~s^{-1}$ in a spherical volume with radius $10$ Mpc $h^{-1}$ around each cluster. Further, we assign SFRs to the simulation halos based on the quenching model described in Eqn. \ref{eqn:quenching_model}. The sample of clusters and the halos in their neighborhood is the same as the one used in previous sections on simulations (\ref{sec:sub_sims}), extended out to $10$ Mpc $h^{-1}$.

\subsubsection{Initial distribution of SFR}
\label{sec:init_colors}

To constrain the quenching timescales of galaxies within the cluster halos it is essential to correctly model the initial distribution of SFRs before they were accreted and became satellites. To model the distribution of SFR of field/central galaxies (i.e. galaxies that have not fallen into larger host halos) we use the Universe Machine simulations \citep{Behroozi:2019kql}. Universe Machine is an empirical model that populates CDM $N$-body simulations with galaxies using an extensive set of observational constraints. It provides galaxy properties like stellar masses, SFRs that have been assigned based on a detailed parameterization of the galaxy-halo connection based on the growth history of each halo.   

In general, the SFR distribution in the Universe at a given redshift is a bimodal function, with its two peaks at high and low SFRs corresponding to star-forming and quiescent galaxies. The fraction of quenched or star-forming galaxies, in turn, is a function of the galaxy stellar mass or alternatively the virial mass of its host dark matter halo. Initial SFRs are therefore assigned to halos based on redshift and maximum circular velocity, $v_{\rm cmax}$, which is a proxy for halo mass. SFRs are assigned to subhalos of the cluster based on their $v_{\rm acc}$, i.e. the $v_{\rm cmax}$ at the time of accretion, and from the distribution of SFR of central galaxies at the accretion redshift, $z_{\rm acc}$.  In detail, we model the distribution of SFR of galaxies as a double Gaussian in bins of $v_{\rm cmax}$ for each redshift,

\begin{equation}
f(SFR_i)= c_1 G(\mu_1, \sigma_1) + c_2 G(\mu_2, \sigma_2),    
\label{eqn:g_dist}
\end{equation}
where $G(\mu,\sigma)$ is a Gaussian with mean $\mu$ and width $\sigma$, $c_1$ and $c_2$ are parameters that control the relative fraction of galaxies in the two Gaussians. The variables $\mu,~\sigma$, and $c$ all depend on redshift and the $v_{\rm cmax}$ of the halo. We use Universe Machine to calibrate these parameters for both central and satellite galaxies at all redshift snapshots between $z=$0--4 available in Universe Machine, in three bins of $v_{\rm cmax}$: [$150 < v_{\rm cmax}<170$, $170 < v_{\rm cmax} < 200$,  $v_{\rm cmax} > 200$] in units of km$~s^{-1}$. Subhalos of the cluster are assigned initial SFRs by randomly drawing from the Gaussian distribution of SFRs described by Eqn. \ref{eqn:g_dist} at $z_{\rm acc}$ for the corresponding $v_{\rm acc}$ bin, where the parameters of the Gaussian are calibrated using central galaxies from Universe Machine.

SFRs are also assigned to all halos outside the virial radius of the cluster out to $10$ Mpc $h^{-1}$ based on their $v_{\rm cmax}$ or $v_{\rm acc}$ (if they are subhalos). They are assigned SFRs from the distribution calibrated from centrals if they are centrals themselves, and from the SFR distribution calibrated from satellites if they are subhalos of other halos.\footnote{We find that the region around clusters can have large number of galaxies that are satellites within larger halos, these satellites tend to have lower SFRs than centrals and their distribution should therefore be drawn from satellite SFR distributions.}

Once  the initial distributions of SFRs are assigned, with galaxies within subhalos of the cluster set to initial states before infall, we evolve the SFRs of all halos based on the quenching model Eqn. \ref{eqn:quenching_model}. 

We summarize our main steps here:
\begin{enumerate}
\item Galaxies are assigned to all $v_{\rm peak}>150 ~\rm km~{s^{-1}}$ halos around cluster-mass halos.   

\item All galaxies are assigned initial SFR based on their $v_{\rm cmax}$ and  redshift by drawing from distributions of $v_{\rm cmax}$--$SFR(z)$ calibrated using Universe Machine.

\item Galaxies in subhalos are assigned SFRs based on the $v_{\rm cmax}$ at their accretion redshift, $z_{\rm acc}$. 

\item Galaxies are evolved with the quenching model from their initial SFRs.
\end{enumerate}

In the following sections we describe the constraints on the quenching model obtained from the location of the steepest slope $r_{\rm smin}$ and the number density profiles of galaxies.

\subsubsection{Constraints from slope-minimum, $r_{\rm smin}$}
\label{sec:constrain_rsmin}

Here we describe the constraints we obtain on the galaxy quenching model from the location of the slope minimum, $r_{\rm smin}$, for the three populations of galaxies with different colors. The free parameters of our model are the delay time, $t_{\rm d}$, and the quenching timescale, $t_{\rm q}$. We divide our parameter space into thirty bins in delay time between $0.1 < t_{\rm d} < 2$ Gyrs and hundred bins in quenching time between $0 < t_{\rm q} < 5$ Gyrs. We evaluate the location of the steepest slopes for mock galaxy populations of different colors at each point in the parameter space and compare it to data. 

The procedure in detail is as follows. Using the time since infall, $t$, provided by the {\sc Rockstar} catalog, we evolve each mock galaxy's initial SFR to the current redshift according to Eqn. \ref{eqn:quenching_model}, for every pair of $t_{\rm q}$ and $t_{\rm d}$ in the parameter space. During the initial period of delay ($t_{\rm d}$), before the onset of exponential quenching, they are evolved according to Eqn. \ref{eqn:field_evn}, following the evolution of isolated galaxies calibrated using Universe Machine.  The evolution timescale, $t_{\rm q,iso}$, varies linearly between $1-3$ Gyrs for the logarithmic velocity range of the subhalos used in this paper. Once the galaxy SFRs have been evolved to the current time (the redshift at which their host cluster is observed), we divide them into red, green, and blue based on the new SFR. The splits in the SFR space to assign simulation galaxies red, green, or blue colors are adjusted to match the color fraction profile outside the splashback radius, by fitting the color fraction profiles from the radial bin at $r=2.64$ Mpc $h^{-1}$ outwards (see \ref{sec:color_ratio} for a detailed description). Following the color split, we compute the slope profile for mock galaxy populations in each color bin, and find the location of the steepest slope, $r_{\rm smin}$. We compare the location of the steepest slope at each point in the plane of $t_{\rm d}$ and $t_{\rm q}$ to the values obtained from data.

\begin{figure}

	\includegraphics[width=0.9\linewidth]{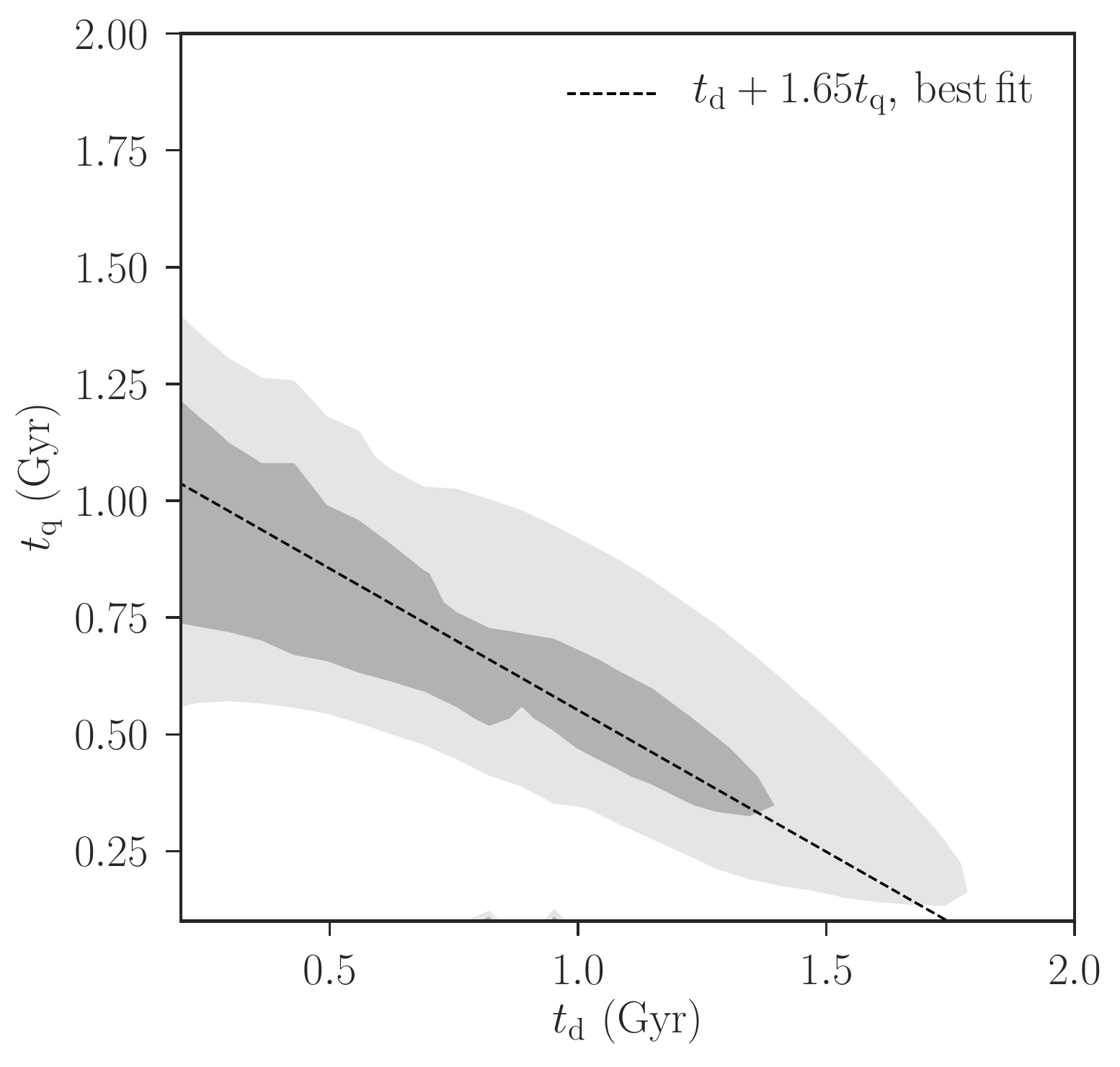}
	\includegraphics[width=0.9\linewidth]{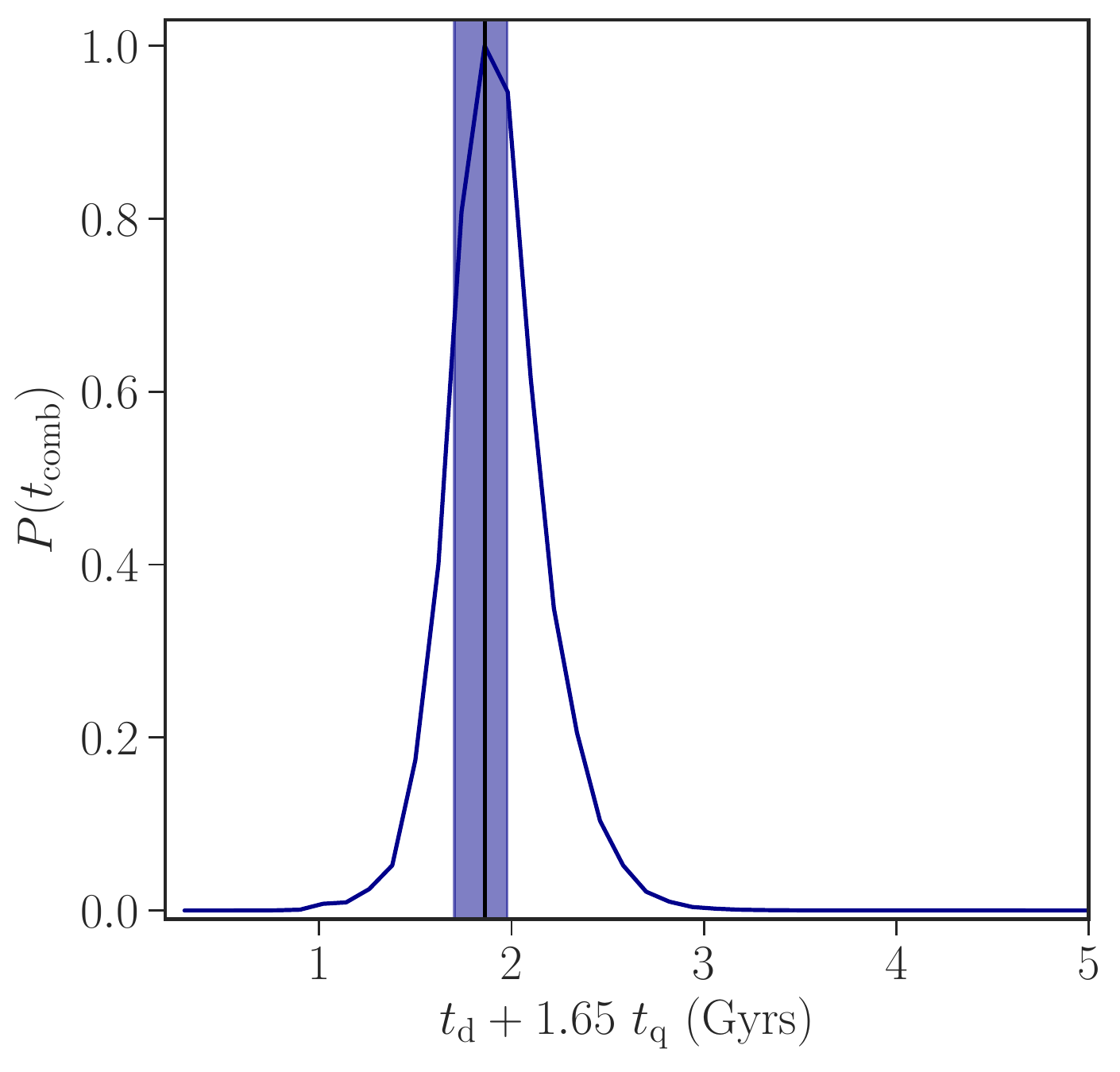}
	\caption{{\it (Top)} Constraints on the quenching model parameters from $r_{\rm smin}$. The darker and the lighter grey contours represent the $1 \, \sigma$ and the $2 \, \sigma$ confidence regions, respectively (see Sec.~\ref{sec:constrain_rsmin}). The black dashed line is the best constrained linear combination of the timescales. {\it (Bottom)} Constraints on the best-constrained linear combination of $t_{\rm q}$ and $t_{\rm d}$ from $r_{\rm smin}$. Blue shaded region shows the $1 \, \sigma$ constraint in the linear combination, $1.85^{+0.12}_{-0.15}$ Gyrs. 
	}
	\label{fig:chi2_rsmin}
\end{figure}

The constraints on the model from the location of $r_{\rm smin}$ are shown in Fig. \ref{fig:chi2_rsmin}. The grey contours in the top panel show the allowed region of the parameter space constrained from the ACT DR5 cluster sample. The dark grey region corresponds to the 68\% confidence interval, which represents the statistical uncertainty obtained  using simulations as described above. The likelihood is calculated assuming a Gaussian form. As expected, there is a degeneracy between the delay time and the quenching timescale. The direction of degeneracy is denoted by the black dashed line in the top panel; short delay times with long $t_{\rm q}$ and long delay times with short $t_{\rm q}$ are both allowed by the data. By extrapolating the black line to $t_{\rm d}= 0$ Gyrs to obtain its intercept on the $y$-axis, it appears that a galaxy takes at least $1.15\pm 0.3$ Gyrs to transition from a star-forming phase to a quenched phase in our model, in the sense that the sum of $t_{\rm q}$ and $t_{\rm d}$ is always greater than $\sim 1.15$ Gyr. 

We find that the location of $r_{\rm smin}$ can be used to best constrain the linear combination $t_{\rm d}+1.65 ~t_{\rm q}$ (bottom panel Fig. \ref{fig:chi2_rsmin}). The best-fit value for this combination is $1.85^{+0.12}_{-0.15}$ Gyrs. Note that $r_{\rm smin}$ is sensitive to the \textit{total} time that a galaxy population spends inside the cluster, particularly before apocenter crossing,  therefore it constrains the linear combination comprised of the sum of the total delay time and a multiple of the $e$-folding time of quenching $t_{\rm q}$, considering the number galaxies with a given SFR reduces by $\sim 65\%$ after first $e$-folding. 

While the location of the minimum gives us an intuitive understanding of the phase-space picture of galaxies that is easy to interpret, there is, theoretically, more information contained in the entire density profile measured from this rich data set.  Referring to Fig.\ref{fig:phase_particles}, $r_{\rm smin}$ gives us the location of the discontinuity between multi-stream and single-stream in phase space, therefore it can tell us about the maximum age of a galaxy population within a halo. On the other hand, the density profile is the integration of the number of galaxies in different streams at each radius; its shape therefore contains information about the radial location of the transition from delay to quenching phases, helping to break the degeneracy between the two timescales. In the next section we use the color ratios to constrain the quenching parameters.

\subsubsection{Constraints from color fraction profile}
\label{sec:color_ratio}

\begin{figure}
    \includegraphics[width=1.0\linewidth]{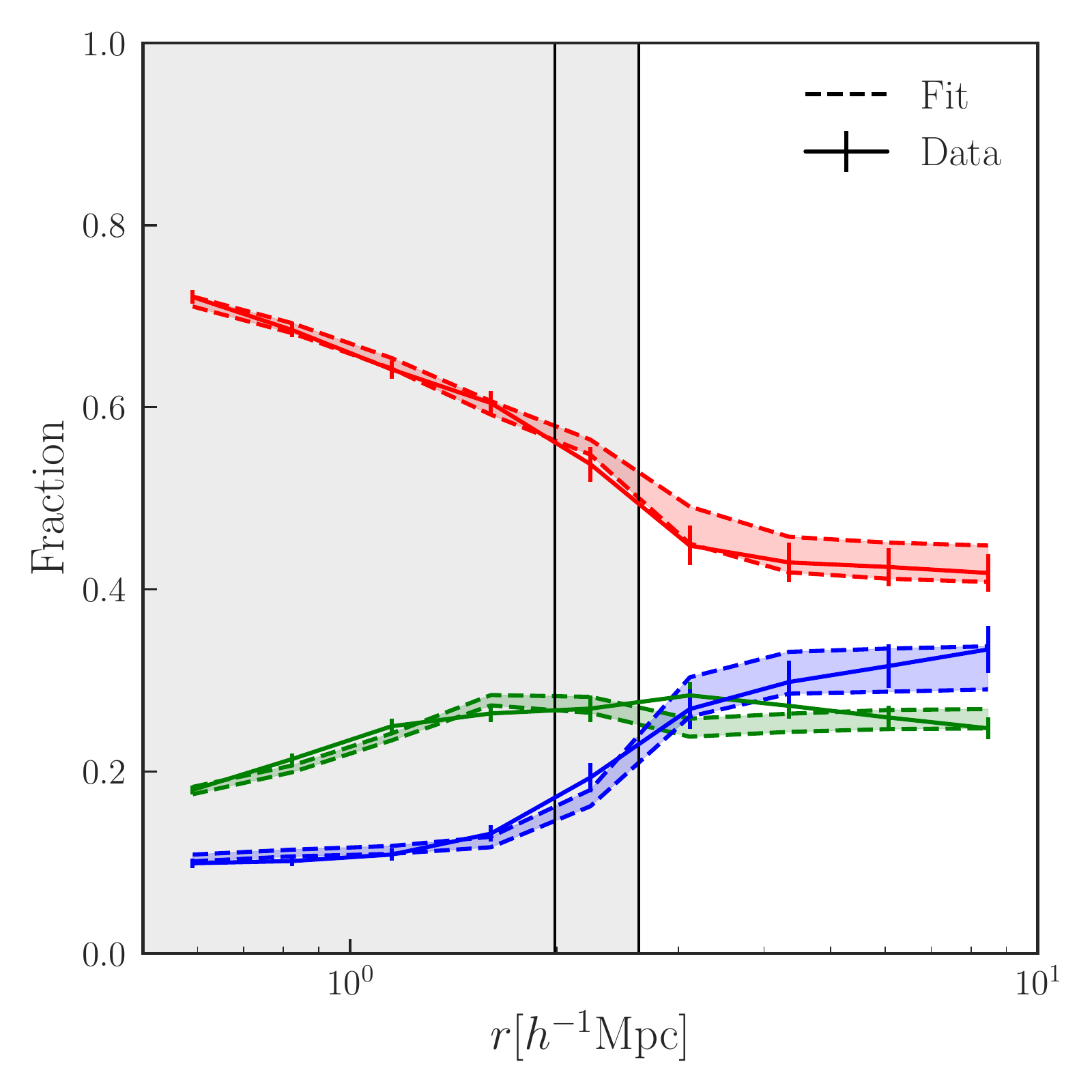}
    \caption{Color fraction - The measured profiles of the fraction of red/green/blue galaxies (solid lines with error bars) as a function of 3-dimensional cluster-centric radius, as defined in Sec.~\ref{fig:color_selection}. The shaded regions in the red, green and blue colors are the 1$\sigma$ confidence intervals from fitting the quenching model to the color ratio. The $\chi^2$ per degree of freedom of the fit is 1.46. The inner (grey shaded) and outer cluster regions are fitted separately. The vertical lines denote the 1$\sigma$ interval of the measured splashback radius for all galaxies in the SZ cluster sample.}
     \label{fig:color_frac_allz}
\end{figure}

To break the degeneracy between $t_{\rm q}$ and $t_{\rm d}$, we explore constraints from the complete number density profile of galaxies in this section. In particular we use the color fraction profile, the ratio of the number of galaxies of a particular color at a given radius to the total number of galaxies at that radius. The solid lines with error bars in Fig.~\ref{fig:color_frac_allz} show the fraction of galaxies in each bin of color around the ACT DR5 cluster sample, as a function of 3D cluster-centric radius. The 3D color fraction profiles and their covariance matrices are retrieved from the MCMC chains of the model fitting described in the Sec.~\ref{sec:splashback_measurement}.

We use the best-fit color fraction curves with $1\sigma$ error bars as input data to constrain the quenching model. We fit our quenching model to nine radial bins between a radius of 0.6 and 10 Mpc $h^{-1}$. Note that we do not use any radial bins smaller than $0.6$ Mpc $h^{-1}$ as the subhalo distributions in the simulations are known to deviate significantly from observed galaxy distribution in this region. In particular, subhalos can be stripped to masses below the halo-finder resolution limit due to tidal stripping or artificial disruption \citep[e.g.][]{vandenBosch:2017ynq}, an effect that is not necessarily reflected in the galaxies. Moreover, blending near cluster centers may lead to an underestimation of the number of galaxies in the innermost bins \citep{Zhang2015}. This effect is particularly important for our analysis as blue galaxies are systematically fainter than red ones \citep{Bell:2003gb, Strateva:2001wt} creating a possible systematic bias in the color ratio due to incompleteness near the center. Although we expect our absolute magnitude cut to mitigate some of this effect, it provides us with additional motivation to use only the radial bins larger than $0.6 ~{\rm Mpc} ~h^{-1}$. We do not expect systematic effects for the innermost galaxies to affect the slope or ratio measurements in the outskirts.

We  constrain our quenching model by comparing the measured fraction of galaxy colors (Fig.~\ref{fig:color_frac_allz}) to that calculated around the clusters in the MDPL2 simulations using the halo number density profiles around them. 
The color fraction profiles shown in Fig.~\ref{fig:color_frac_allz} span a large range that encompasses the virialized region and the region where halos are still falling into the cluster. Bearing in mind that the infalling halos outside the clusters could themselves be part of larger structures like group-mass halos whose quenching timescales can, in principle, differ from the timescales within clusters, we treat the inner and outer regions around the cluster separately. We denote the radial bins outside the splashback boundary of the cluster sample with $r\ge 2.64$ Mpc $h^{-1}$ as the ``outer'' region of the cluster and the radial bins within $2.64$ Mpc $h^{-1}$ as the inner region. We allow the quenching timescale of satellites outside the cluster to be different from that inside it. 

For clarity we discuss our results from the outer and inner regions separately below.

\begin{figure}
	\includegraphics[width=1.0\linewidth]{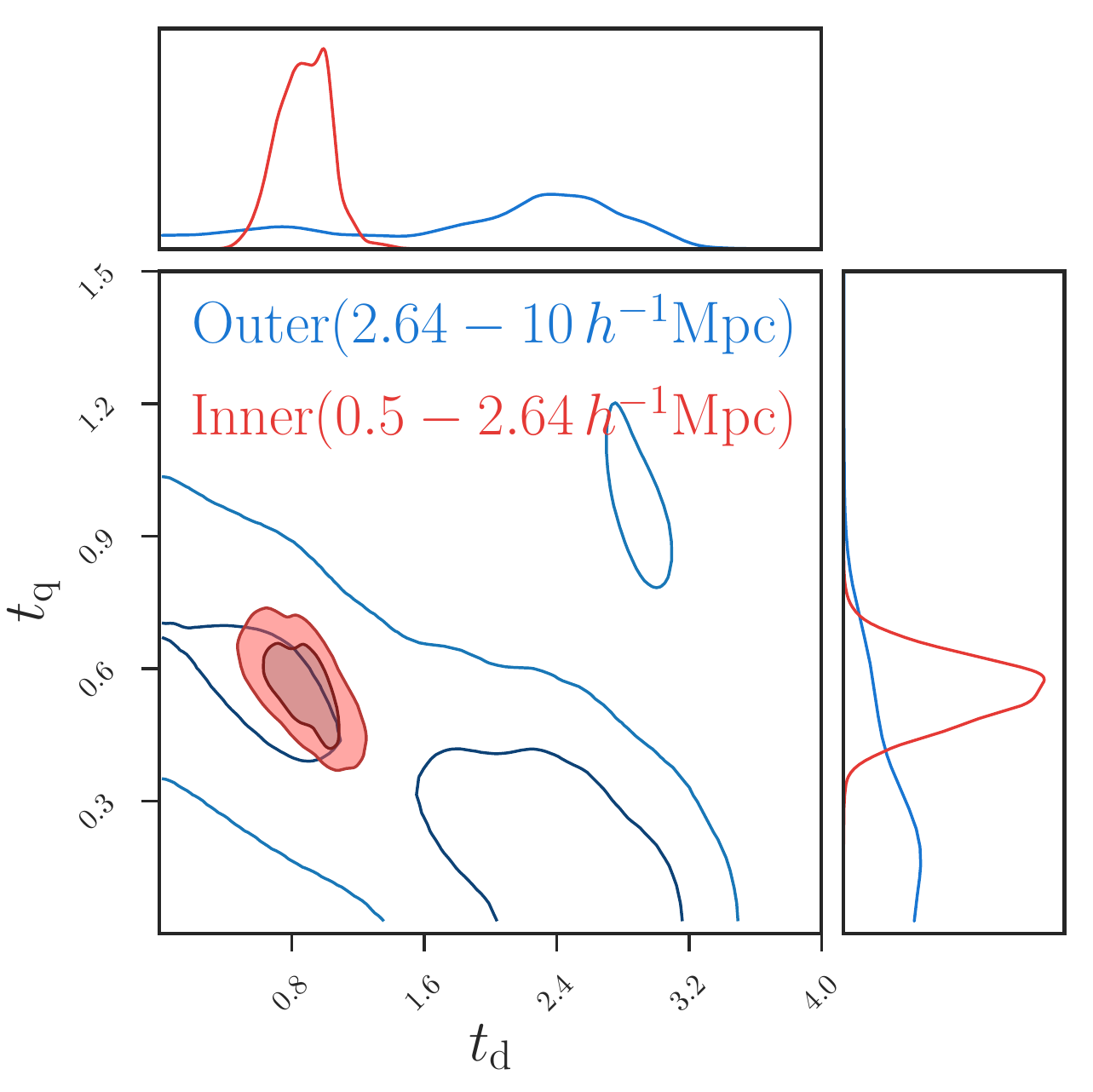}
	
	\caption{Constraints on quenching timescales from 3D color fraction profiles. (Red shaded) The $1 \, \sigma$ and the $2 \, \sigma$ contours from fitting the color fraction profile between 0.6 - 2.64 Mpc $h^{-1}$ (``inner'' profile). \textit{The red shaded region corresponds to quenching timescales within the clusters from ACT DR5.}
	(Blue lines) The constraints on parameters using the color fraction profile of the ``outer'' region beyond 2.64 Mpc $h^{-1}$.
	The darker and the lighter blue contours correspond to the 68\% and 95\% confidence region respectively. These correspond to quenching timescales within halos \textit{outside} the virial radius of the cluster.
	}
	\label{fig:chi2_inner_outer}
\end{figure}

\textit{Constraints from the outer regions of clusters}: To map from the observed color space to the SFR space in simulations, we only use the color fraction in the radial bins outside the clusters. We define our boundary between red,green, and blue galaxies in SFR space based on the relative density of red, green, and blue galaxies between $2.64-10 ~{\rm Mpc} ~h^{-1}$, as this region does not evolve with the quenching timescales within the cluster. 

We use two parameters, $f_{\rm rg}$ and $f_{\rm gb}$ to specify the splits in the  SFR space to assign simulation halos to red, green, and blue color bins. The value of SFR that separates red from green, and green from blue are described as follows: 

\begin{align}
  SFR_{\rm rg}=f_{\rm rg} SFR_{\rm r} + (1-f_{\rm rg}) SFR_{\rm g}  \,\,\,\,\, &\nonumber\\
  SFR_{\rm gb}=f_{\rm gb} SFR_{\rm g} + (1-f_{\rm gb}) SFR_{\rm g},  \,\,\,\,\, &
\end{align}
where $SFR_{\rm r}$ and $SFR_{\rm b}$ correspond to the location of the two peaks at low and high SFR in the bimodal SFR distribution respectively.  $SFR_{\rm g}$ is the location of the minimum separating the two peaks, often referred to as the green valley. The splitting points are described with respect to peaks in each individual $v_{\rm cmax}$ bin.  

Reviewing Sec \ref{sec:init_colors}, the SFRs of central halos at $r>2.64 ~{\rm Mpc}~ h^{-1}$ are assigned based on their current $v_{\rm cmax}$. However, subhalos in this region that are within the virial radius of more massive halos, have an accretion time associated with them that corresponds to the redshift at which they crossed into the virial radius of their current host. We assign initial SFRs to these objects based on the $v_{\rm cmax}$ at the time of accretion onto their host halos and use the quenching model to evolve them to the observed redshift. This allows us to infer quenching timescales associated with groups outside and around the ACT DR5 clusters. 

The profile in the outer region therefore allows us to fix the color--SFR mapping and additionally constrain any quenching timescales associated with groups outside the cluster. We refer to the quenching parameters in the outer region as $t_{\rm d,out}$ and $t_{\rm q,out}$. 

\begin{figure*}
    \includegraphics[width=1.\linewidth]{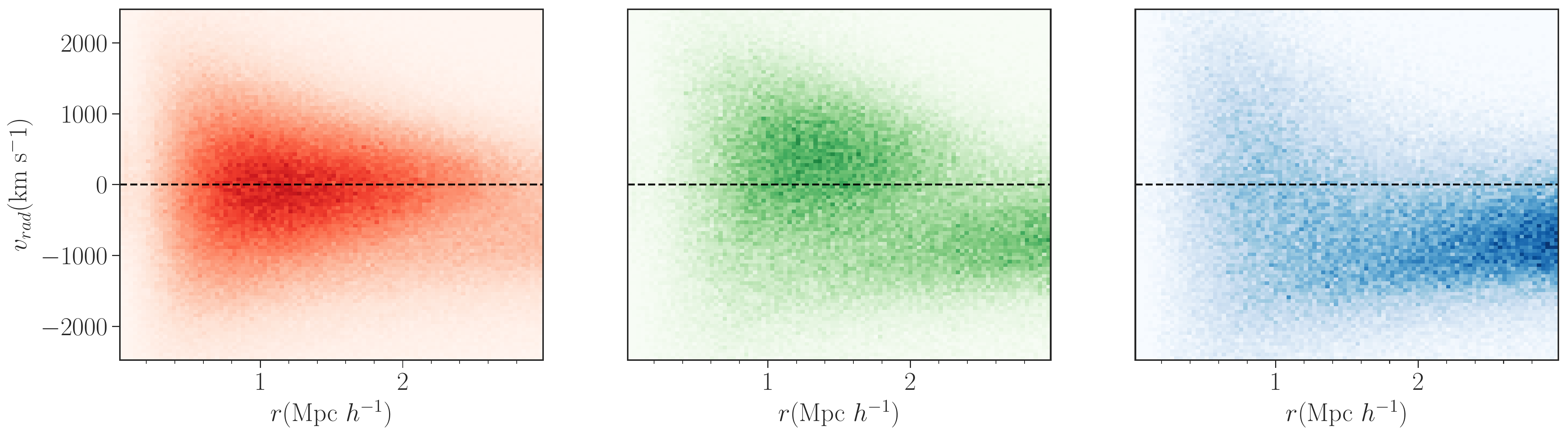}
    \caption{The inferred distribution of galaxies of different colors in phase space from the best-fit parameters of the quenching timescales obtained from data. The phase-space diagram has been generated using simulations halos that have been assigned colors based on the quenching model. The three panels correspond to red, green, and blue galaxies from left to right. The red galaxies appear to be mostly concentrated in the virialized, multi-streaming region while the blue galaxies are still infalling. The green galaxy population is a combination of infalling galaxies and galaxies that have crossed over to a second stream in phase space but not reached apocenter. (see also Fig. \ref{fig:phase_particles})}.
    \label{fig:best_fit_phase}
\end{figure*}

We fit the outer color fraction profile by running MCMC chains assuming a Gaussian likelihood (see Sec.~\ref{sec:measurement}). We vary the four model parameters, $f_{\rm rg}$, $f_{\rm gb}$, $t_{\rm q,out}$ and $t_{\rm d,out}$, split the galaxy SFRs in simulations into red, green, and blue based on $f_{\rm rg}$, $f_{\rm gb}$ and compute the color fraction profile as a function of the cluster-centric radius. The color fraction profiles in the simulation are then compared to those measured from data (Fig.~\ref{fig:color_frac_allz}) to derive the constraints on the model parameters. 

The blue contours in Fig.~\ref{fig:chi2_inner_outer} show the resultant constraints on the quenching parameters from the outer profiles. The 68\% confidence interval of the delay time, $t_{\rm d,out}$, is 0.8-2.7 Gyrs and $t_{\rm q,out}<0.7$ Gyr at 95\% confidence level (blue contours). We emphasize that these quenching parameters correspond to timescales within subhalo hosts outside the cluster, in a sense they correspond to pre-processing timescales within groups before infall. The typical mass of a  subhalo-host in the infall region is $10^{12.5} M_\odot h^{-1}$. This result is consistent with previous studies for quenching timescales around group-sized halos, including \cite{Wetzel:2012nn} which reports a delay time of 2--4 Gyrs and an upper limit for the quenching timescale of 0.8 Gyrs. 

\begin{figure}
    \includegraphics[width=0.9\linewidth]{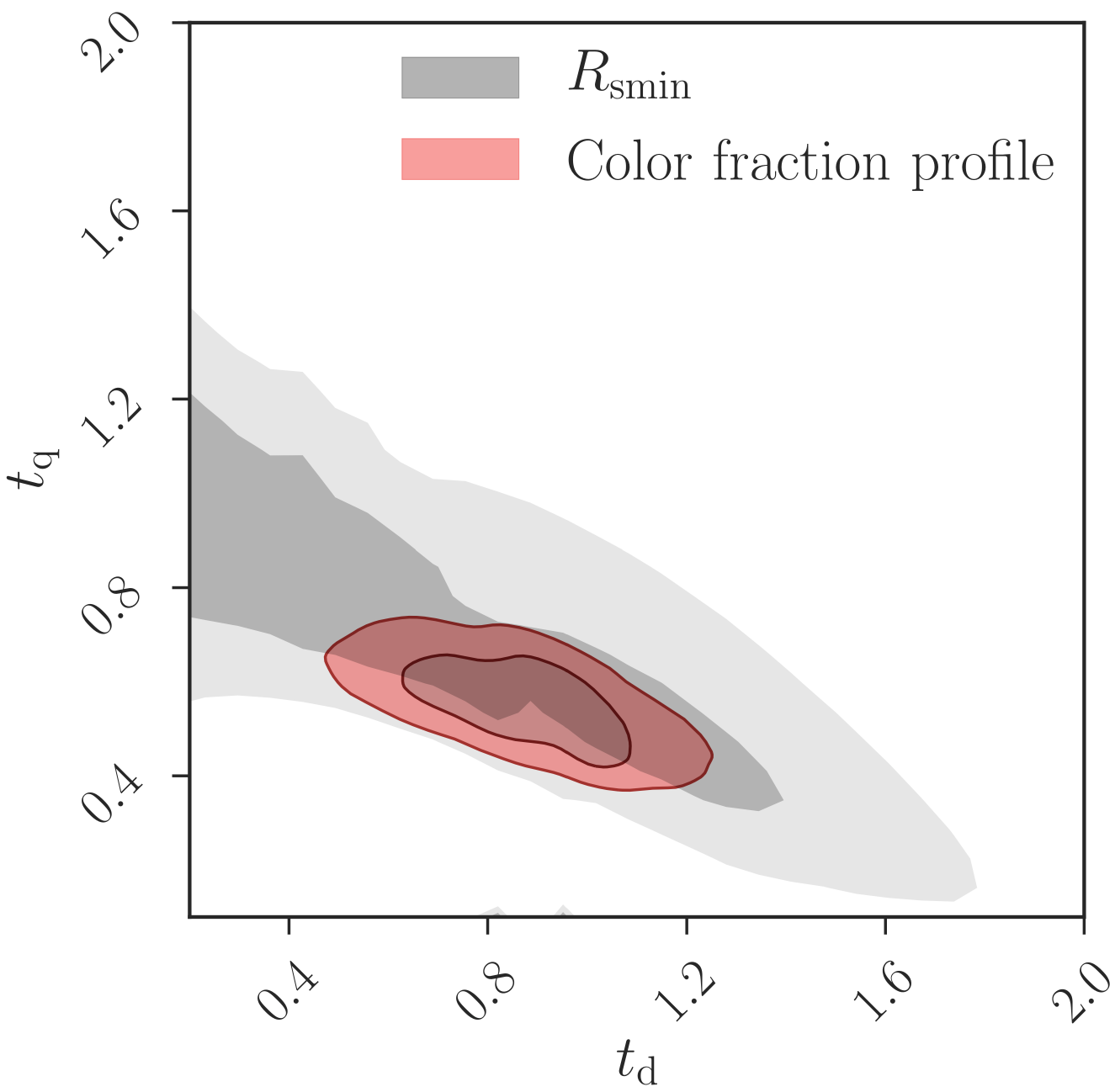}
    \caption{ Comparison between constraints on $t_{\rm q}$ and $t_{\rm d}$ from two techniques: the location of $r_{\rm smin}$ and 3D color fraction profiles.
    The grey bands correspond to the $1\sigma$ and $2\sigma$ constraints from $r_{\rm smin}$. The red shaded contour corresponds to constraints obtained using the entire inner profile (same as Fig. \ref{fig:chi2_inner_outer}).
    }
    \label{fig:contour_comp}
\end{figure}

\textit{Constraints from the inner profiles of clusters}: The analysis on the inner region is performed in a similar manner as the outer region. We adopt the splitting locations $f_{\rm rg}$ and $f_{\rm gb}$, constrained by the outer region as priors for the color splits. As before, at every step in the MCMC chain we vary four parameters, the two splitting locations and the quenching parameters in the inner cluster, $t_{\rm q,in}$ and $t_{\rm d,in}$, where `in' denotes the inner region.

The final constraint on the quenching parameters for the interior of the cluster is shown by the red contours in Fig. \ref{fig:chi2_inner_outer}. The best-fit value for the exponential quenching timescale, $t_{\rm q,in}$, is $0.6^{+0.1}_{-0.1}$ and for the delay time, $t_{\rm d,in}$, is $1.0^{+0.1}_{-0.3}$. While the quenching timescale, $t_{\rm q,in}$, is consistent with that in \citet{Wetzel:2012nn}, we find a significantly smaller the delay time, $t_{\rm d,in}$,  than \citet{Wetzel:2012nn}. It appears that the onset of exponential quenching happens at an earlier time after infall in the massive clusters that we observe compared to the lower mass, group-sized objects explored in \citet{Wetzel:2012nn}. This can be attributed to the fact that the density of gas and dark matter in lower mass objects can be significantly smaller compared to massive clusters used in this study, allowing them to survive over longer times and more orbits before quenching begins. 

In Fig. \ref{fig:best_fit_phase} we illustrate the likely distribution of the galaxies in the phase space of dark matter halos using the best-fit quenching parameters. We generate the distribution of galaxies in the $r-v_{\rm rad}$ plane using the simulated galaxies and separating them in color space. The color indicates the fraction of galaxies in each sample that are in a given pixel. Our quenching parameters indicate that the blue galaxies are primarily on first infall, having only recently crossed pericenter, whereas the red and green galaxies are in the virialized region of the cluster. We note that previous work like \citet{Adhikari:2018nxy}, \citet{Oman_2016} and \citet{Orsi:2017ggf} that have used spectroscopic data to infer dynamical properties of galaxies also conclude, similar to our findings, that blue galaxies have been accreted onto their host clusters recently and live mostly in the infall stream. Our method, combined with the inference drawn from the location $r_{\rm smin}$, implies further that they are past pericenter in their orbits. We do not compare the values of the timescales in these works here due to different cluster and color selection choices made in the papers mentioned above.

\subsubsection{Discussion on quenching constraints}

Fig. \ref{fig:contour_comp} shows the comparison between the constraints obtained on the quenching timescales within clusters from the two different methods. The grey and red shaded regions show the  constraints from $r_{\rm smin}$ and the 3D color ratio profiles respectively. The two results are consistent with each other statistically. The color fraction profiles give relatively tighter constraints due to the greater amount of information contained in them. While the relative locations of $r_{\rm smin}$ allow for longer quenching times and shorter delays, the color fraction profile breaks the degeneracy to a large extent, preferring a delay time that is close to $1~\rm Gyr$. Comparing with subhalo accretion scales from simulations, this result implies that the onset of exponential quenching begins close to the pericenter of a cluster, as it takes a subhalo approximately a gigayear to reach pericentric passage from the virial radius of a cluster. 

We note some caveats that must be considered with regard to the methods used in this paper. Our simple quenching model with two free parameters does not include all modeling uncertainties that may be present in the galaxy--halo connection. For example, we do not include scatter in the abundance matching relationship and we also do not include scatter in the relation between the SFR and $v_{\rm cmax}$; both of these can broaden the error contours shown in \ref{fig:color_splashback_data}. Another theoretical uncertainty not included in this work is the modeling of orphan galaxies in simulations; these are galaxies that are expected to live within destroyed subhalos in the simulation. Subhalos suffer enhanced disruption compared to galaxies as they are more extended and feel stronger tidal forces, and also suffer from artificial disruption due to resolution effects; not accounting for galaxies within such subhalos can significantly change the cluster profiles. Therefore we caution against a direct comparison of the slope and number density profile from the innermost region of clusters in simulations to the slopes observed in galaxies of real clusters.
Both of these effects are stronger near cluster centers and we have mitigated them by considering galaxy profiles outside the very center of the cluster by using only the region, $r>0.6$ Mpc $h^{-1}$ in our analysis; we also use relative color fractions for comparison instead of the absolute values of the number density at each radius. We also note that while the difference in the maximum time spent within the cluster by different galaxy populations can explain the relative location of splashback  between galaxies with different colors, it does not necessarily explain the complete shape of the density profile. Comparison between Fig. \ref{fig:splashback_subhalos} and \ref{fig:color_splashback_data} clearly demonstrates this point. While the location of splashback for green galaxies agrees with simulations, these galaxies show significantly shallower slopes in the inner region in observation compared to simulations.

Furthermore, in this work we do not explicitly account for differences in quenching timescales between galaxies that are on different types of orbits. In principle, galaxies on radial orbits that form the inner regions of the cluster can have different quenching timescales from those that are on tangential orbits, as they probe different cluster environments. We note that while our model does not explicitly include orbital parameters, infall times themselves correlate with pericenter distances. We defer a detailed quantitative analysis of this effect to future studies. With regard to the location of $r_{\rm rsmin}$, we note that while the radial profile can be significantly different for galaxies in tangential orbits compared to those on radial orbits, we find that the splashback boundary for these populations does not move significantly, therefore differences in orbital angular momentum cannot alone explain the behavior of $r_{\rm smin}$ observed in data (see Appendix).

In this paper we have used CDM-only simulations to map the distribution of observed galaxies around clusters to satellites in dark matter halos. However, in principle a more detailed study of the specific effects of ram pressure or harassment that lead to quenching can be conducted using controlled hydrodynamical simulations. We note that some of the issues like disruption of satellites near cluster centers are less important in hydrodynamical simulations as they simulate galaxies directly, however the exact timescales involved in star-formation evolution can have significant uncertainties and depend strongly on modeling of feedback mechanisms. The timescales obtained from this analysis appear to be consistent with that found in hydrodynamical simulations for massive clusters in \cite{Bahe:2014eea}, however the delay time in our work is smaller compared to \cite{Rhee_2020}. We note that the cluster sample in our work is significantly more massive and at a higher mean redshift compared to both the studies quoted above, and a detailed comparison with hydrodynamical simulations with the mass and redshift distribution matched to the cluster sample is a useful direction for future work. 

\section{Conclusion}
\label{sec:conclusion}

We study the distribution of galaxies around  SZ-selected galaxy clusters as a function of galaxy color. The galaxy sample is derived from the DES Y3 gold galaxy catalog, and the SZ-selected clusters are identified by AdvACT and will be published as a part of ACT DR5 \cite{ACTcluster_inprep}. The number density profiles encode important information about the dynamical history of galaxies in these massive systems; in this paper we study them in the context of the splashback radius $r_{\rm sp}$. 
We invoke a new parameter $r_{\rm smin}$ that generalizes $r_{\rm sp}$, as it represents the location of the steepest slope even for galaxies that do not reach splashback.  This occurs most distinctly for blue galaxies that are rapidly quenched and become red before reaching apocenter. 

We demonstrate that even in the absence of any velocity information, the profiles can be used to extract time evolution information for segments of the galaxy population. The time elapsed between crossing into the cluster and reaching $r_{\rm smin}$ traces the maximum time members of a galaxy population have been inside a cluster,  in a model-independent way. Populations of galaxies with different SFRs can be mapped to halos that have entered clusters at different times.  This allows us to estimate the timescale(s) associated with galaxy quenching. 

Using only photometric data and the  spatial distribution of galaxies, we are thus able to map galaxies onto different regions of the 3D phase space of dark matter halos. Just like the splashback radius traces the boundary between the virialized, multi-streaming region and the infall region, $r_{\rm smin}$ traces the discontinuity in the phase-space distributions even for galaxies that  do not reach  splashback. While the galaxy distribution in the inner regions of massive clusters can be significantly sensitive to baryonic physics, the location of $r_{\rm smin}$ is more robust. Further, the information in the full density profiles, excluding  the inner-most regions  but including  the slope-minimum, provides more detailed constraints on the parameters for quenching. 

We use CDM-only simulations to create mock galaxy catalogs that reproduce the observed galaxy distributions of different colors and study their density distribution and slope profiles. We use a simple quenching model to estimate the quenching timescales for galaxies in cluster-mass halos corresponding to our SZ sample. Our principal empirical findings are: 
\begin{enumerate}
\item The location of the splashback radius for the SZ-selected clusters in ACT DR5, measured using the complete galaxy sample is $2.4^{+0.3}_{-0.4}$ Mpc $h^{-1}$. This is consistent with theoretical expectations from $N$-body simulations. 

\item The shape of the density profile of galaxies differ significantly as a function of galaxy color. The observed density profiles show evidence of shifts of $r_{\rm smin}$, the location of the steepest slope; in particular the blue galaxies appear to show a weak minimum at a smaller radius compared to red and green galaxies (see Fig \ref{fig:color_splashback_data}).

\item By comparing  $r_{\rm smin}$ estimated for galaxies with simulated subhalos accreted at different times, we find that blue galaxies largely live in the infall stream, have not reached splashback, and are likely to have been accreted between 1.1 and 1.5 Gyrs ago. Red galaxies show a sharp slope minimum at $2.2 ~{\rm Mpc}~ h^{-1}$ and have been in the halo for more than $3.2 ~\rm Gyrs$ on average, while green galaxies have been in the halo for at least  $2.3 ~\rm Gyrs$ on average and have nearly reached their apocenter (see Fig. \ref{fig:splashback_subhalos}). 

\item We use the entire color fraction profile of galaxies between $0.5 {\rm ~Mpc}~h^{-1}  $ to $2.6 ~{\rm Mpc}~h^{-1}$ to obtain constraints on our quenching parameters (see Fig. \ref{fig:chi2_inner_outer}). We find that the delay time is $t_{\rm d}=1.0^{+0.1}_{-0.3} ~ \rm Gyr$ and the exponential quenching timescale is $t_{\rm q}=0.6^{+0.1}_{-0.1} ~ \rm Gyr$. The constraints obtained using the profiles agree with the timescales obtained from the location of the slope minimum. The color fraction profiles thus imply a short quenching time and a longer delay time that is comparable to the pericenter crossing time of the cluster. 

\item Using color ratio profiles in the region outside the virial radius of the cluster, we find that the total time required for quenching of satellite galaxies in this region is longer than  inside the cluster, as expected since these galaxies are likely to occupy group-sized halos. The delay time is constrained to be in the range $0.8 < t_{\rm d} < 2.7$ Gyrs and the exponential quenching time  is constrained to $t_{\rm q} < 0.7 ~{\rm Gyr}$ (see Fig \ref{fig:chi2_inner_outer}).

\end{enumerate}

We note that the error bars quoted in this paper are statistical and derived from a comparison with $\Lambda$CDM simulations without baryons. The simulations can introduce systematic biases particularly in the innermost radial bins, as discussed in Section 4.3.4. Because we focus mainly on the region near the outskirts of galaxy clusters, these effects are not expected to be severe. We do not include any baryonic effects on subhalo orbits and evolution. These can impact the galaxy--halo connection, and a detailed comparison of these results with the density and slope profiles of galaxies in hydrodynamic simulations is a promising direction to explore in the future. 

The splashback radius is a robust lengthscale within dark matter halos that can be used to understand the evolution of non-linear, virialized structures in the Universe and the evolution of galaxies within them. We emphasize that while the location of the minimum of the density slope appears to be a function of galaxy color, the splashback radius, i.e., the location of the first apocenter of galaxy orbits, is not. The movement of the location of the minimum is due to the movement of the phase-space boundary between a single stream and multi-stream region. 

We note that the location of the steepest slope for red galaxies is consistent with the location of $r_{\rm smin}$ for the entire galaxy sample, implying that the total galaxy sample provides a robust estimate of the location of splashback.
This location, we expect, should also be consistent with the splashback radius measured from dark matter particles itself for the samples in our data --- specifically  galaxies that are not experiencing significant dynamical friction \citep{Adhikari:2016gjw, More:2016vgs, Baxter:2017csy, Chang:2017hjt}. We also note that while the location of $r_{\rm smin}$ for red galaxies is the same as that of the total galaxy sample, the feature itself is much sharper and deeper in the latter case. This is further evidence that the slope minimum (splashback radius in this case) traces the cluster boundary, where a caustic-like feature is expected from galaxies/particles on turn-around. The sharpening of the feature is similar to the sharpening observed in \citet{Adhikari:2014lna}, Fig. 1, wherein splashback was measured by sub-selecting particles with low radial velocities, tracing the caustic at turn-around or apocenter more clearly.

In this paper we have mostly focused on the most massive clusters and on the mean trend for clusters in a range of redshifts. We find that blue galaxies trace recent accretion. In particular, they infer that blue galaxies should be present mostly in the infall stream. In theory, halo accretion history can differ as a function of redshift, for example a similar mass halo at two different redshifts will have different growth rates; studying star-formation evolution as a function of redshift using the distribution of galaxies in phase space is an intriguing possibility in the future. High redshift spectroscopic studies of X-ray detected clusters like \citet{Willis:2020ixk} and \citet{Miller:2018xwl} also provide the opportunity to study clusters in their early formation stages.

The analysis in this paper can naturally also be extended to lower-mass clusters and groups that can be found using optical selection or using low mass $X$-ray and SZ clusters from upcoming surveys like  eRosita \citep{Predehl:2010vx, Pillepich:2011zz}, Simons Observatory \citep{Ade:2018sbj}, and CMB-S4 \citep{Abazajian:2019eic}. Further, we can study quenching timescales as a function of cluster properties; existing DES data can already be used to study quenching as a function of properties of the host cluster, e.g. stellar mass \citep{Palmese:2019lkh}, redshift, and other tracers of history.  We defer this natural extension to future work. In the immediate future, DES Year 6 data will allow us to probe fainter galaxies with lower stellar mass than our current sample around clusters that are within the DES footprint. This will enable a study of quenching jointly as a function of galaxy stellar mass and SFR. Ongoing and future galaxy surveys should provide an unprecedented wealth of data that can be used in novel ways, like studying the galaxy distribution in the light of the splashback radius, to understand more deeply the connection between galaxies and their dark matter halos.

\begin{acknowledgments}

We thank Arka Banerjee, Peter Behroozi, Benedikt Diemer and Andrey Kravtsov for illuminating discussions and reviewing early drafts of the paper. This work was supported in part by the U.S. Department of Energy contract to SLAC no. DE-AC02- 76SF00515 and made use of computational resources at SLAC National Accelerator Laboratory, a U.S.\ Department of Energy Office and the Sherlock cluster at the Stanford Research Computing Center (SRCC); the authors are thankful for the support of the SLAC and SRCC computing teams. The CosmoSim database used in this paper is a service by the
Leibniz-Institute for Astrophysics Potsdam (AIP). The MultiDark
database was developed in cooperation with the Spanish MultiDark
Consolider Project CSD2009-00064. JPH acknowledges funding for SZ cluster studies from NSF grant number AST-1615657. NS acknowledges support from NSF grant numbers AST-1513618 and AST-1907657. This paper has gone through internal review by the DES collaboration and the ACT collaboration.

Funding for the DES Projects has been provided by the DOE and NSF(USA), MEC/MICINN/MINECO(Spain), STFC(UK), HEFCE(UK). NCSA(UIUC), KICP(U. Chicago), CCAPP(Ohio State), 
MIFPA(Texas A\&M), CNPQ, FAPERJ, FINEP (Brazil), DFG(Germany) and the Collaborating Institutions in the Dark Energy Survey.

The Collaborating Institutions are Argonne Lab, UC Santa Cruz, University of Cambridge, CIEMAT-Madrid, University of Chicago, University College London, 
DES-Brazil Consortium, University of Edinburgh, ETH Z{\"u}rich, Fermilab, University of Illinois, ICE (IEEC-CSIC), IFAE Barcelona, Lawrence Berkeley Lab, 
LMU M{\"u}nchen and the associated Excellence Cluster Universe, University of Michigan, NFS's NOIRLab, University of Nottingham, Ohio State University, University of 
Pennsylvania, University of Portsmouth, SLAC National Lab, Stanford University, University of Sussex, Texas A\&M University, and the OzDES Membership Consortium.

Based in part on observations at Cerro Tololo Inter-American Observatory at NSF’s NOIRLab (NOIRLab Prop. ID 2012B-0001; PI: J. Frieman), which is managed by the Association of Universities for Research in Astronomy (AURA) under a cooperative agreement with the National Science Foundation.

The DES Data Management System is supported by the NSF under Grant Numbers AST-1138766 and AST-1536171. 
The DES participants from Spanish institutions are partially supported by MICINN under grants ESP2017-89838, PGC2018-094773, PGC2018-102021, SEV-2016-0588, SEV-2016-0597, and MDM-2015-0509, some of which include ERDF funds from the European Union. IFAE is partially funded by the CERCA program of the Generalitat de Catalunya.
Research leading to these results has received funding from the European Research
Council under the European Union's Seventh Framework Program (FP7/2007-2013) including ERC grant agreements 240672, 291329, and 306478.
We  acknowledge support from the Brazilian Instituto Nacional de Ci\^encia
e Tecnologia (INCT) do e-Universo (CNPq grant 465376/2014-2).

This manuscript has been authored by Fermi Research Alliance, LLC under Contract No. DE-AC02-07CH11359 with the U.S. Department of Energy, Office of Science, Office of High Energy Physics. 

The ACT project is supported by the U.S. National Science Foundation through awards AST-1440226, AST-0965625 and AST-0408698, as well as awards PHY-1214379 and PHY-0855887. Funding was also provided by Princeton University, the University of Pennsylvania, and a Canada Foundation for Innovation (CFI) award to UBC. ACT operates in the Parque Astron\'{o}mico Atacama in northern Chile under the auspices of the Comisi\'{o}n Nacional de Investigaci\'{o}n Cient\'{i}fica y Tecnol\'{o}gica de Chile (CONICYT). Computations were performed on the GPC supercomputer at the SciNet HPC Consortium and on the hippo cluster at the University of KwaZulu-Natal. SciNet is funded by the CFI under the auspices of Compute Canada, the Government of Ontario, the Ontario Research Fund - Research Excellence; and the University of Toronto. The development of multichroic detectors and lenses was supported by NASA grants NNX13AE56G and NNX14AB58G.

\end{acknowledgments}

\bibliography{references}

\appendix

\label{appendixa}

\section{Splashback as a function of orbital angular momentum of particle orbits}

One possible explanation for the movement of the measured minimum of the density profile may be related to the distribution of orbits of galaxies. It is possible that galaxies that are on tangential orbits quench more slowly than galaxies that are on radial orbits. If tangential and radial orbits reach apocenters at different locations on average, we may expect that the movement of splashback between red, green, and blue galaxies may be explained by the difference in orbital histories. To check this hypothesis we use simulations to study the splashback radius of particles with different pericenters. 
\begin{figure*}
    \centering
    \includegraphics[scale=0.4]{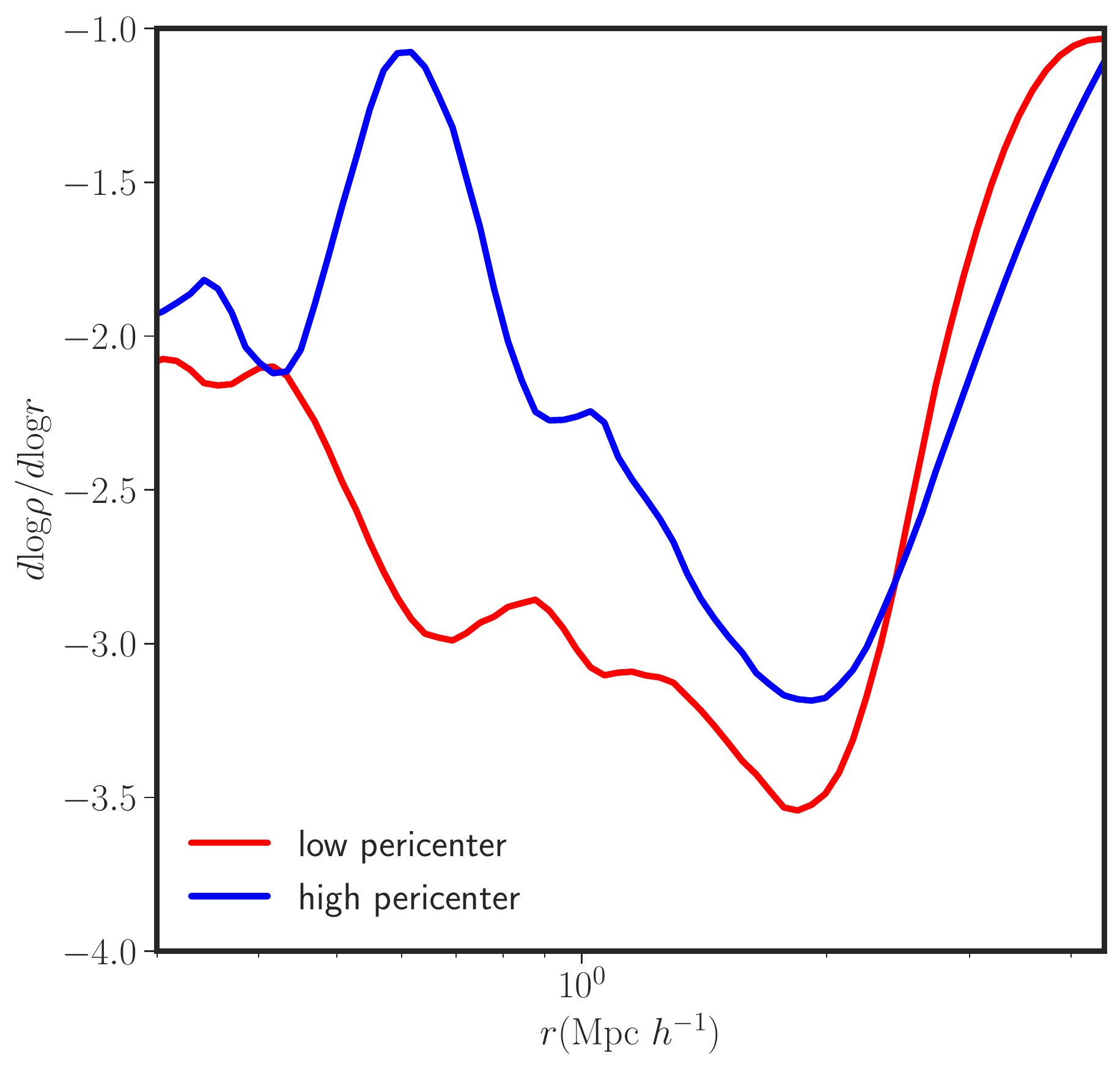}
    \caption{The slope of the density as a function of radius for particles that have been assigned colors based on their different orbital pericenters. the blue curve corresponds to particles that have on average high pericenters and red curve to particles that have low pericenters inside the halo. The location of the minimum of the slope remains unchanged in this toy model. }
    \label{fig:tan_rad}
\end{figure*}

We track the particles orbits from the time they cross into $4 {~\rm Mpc}~ h^{-1}$ comoving distance from the cluster center and find the time and location of their first pericentric passage. Orbits with low pericenters correspond to particles that come in on radial orbits, while those with high pericenters come in with high angular momentum in tangential orbits. We randomly assign colors to all particles within $4 {~\rm Mpc} ~h^{-1}$. We assign particles that have pericenter below $0.6$ Mpc $h^{-1}$ random colors with a red fraction of $0.8$ and particles that have pericenters larger than $0.6$ Mpc $h^{-1}$ a red fraction of $0.4$. In Fig. \ref{fig:tan_rad} we show the logarithmic slope of the density profiles of particles that have been assigned red colors and those that have been assigned blue colors. We find that the location of the outer minimum does not change for the two samples. We find the same result if we alter the pericenter limit of $0.6$ Mpc  $h^{-1}$ to a different value. 

Therefore we conclude that the orbital differences in particles do not affect the splashback radius to a large radius, however the inner density and slopes of particles can be significantly affected by it. 

\end{document}